\documentclass[usenatbib]{mn2e}
\usepackage{natbib}
\usepackage{amssymb}
\usepackage{epsfig}
\usepackage{rotating} 
\usepackage{aas_macros}
\newcommand{\Lx}{\ensuremath{L_{\mathrm{X}}}}

\newcommand{\Msol}{\ensuremath{\mathrm{M_{\odot}}}}

\newcommand{\Dv}{\ensuremath{\Delta_{\mathrm{v}}}}
\newcommand{\rt}{\ensuremath{R_{\mathrm{200}}}}
\newcommand{\rc}{\ensuremath{r_{\mathrm{c}}}}
\newcommand{\rd}{\ensuremath{r_{\mathrm{d}}}}
\newcommand{\Zsol}{\ensuremath{\mathrm{Z_{\odot}}}}
\newcommand{\fgas}{\ensuremath{f_{\mathrm{gas}}}}
\newcommand{\Mgas}{\ensuremath{M_{\mathrm{gas}}}}
\newcommand{\Om}{\ensuremath{\Omega_{\mathrm{m}}}}
\newcommand{\OM}{\ensuremath{\Omega_{\mathrm{M}}}}

\newcommand{\eg}{{\it e.g.\ }}
\newcommand{\egc}{{\it e.g.}}  % Citation version; no space afterwards
\newcommand{\etal}{{\it et al.\thinspace}}

\newcommand{\cf}{{\it c.f.\ }}
\newcommand{\ie}{{\it i.e.\ }}

\newcommand{\Chandra}{\emph{Chandra}}

\newcommand{\ROSAT}{\emph{ROSAT}}
\newcommand{\XMM}{\emph{XMM-Newton}}

\newcommand{\MEKAL}{\textsc{MeKaL}}

\newcommand{\chisq}{\ensuremath{\chi^2}}

\newcommand{\gta}{\,\rlap{\raise 0.4ex\hbox{$>$}}{\lower 0.6ex\hbox{$\sim$}}\,}  
\newcommand{\lta}{\,\rlap{\raise 0.4ex\hbox{$<$}}{\lower 0.6ex\hbox{$\sim$}}\,}  

\newcommand{\nm}{\mbox{\ensuremath{\mathrm{~\nm}}}}

\newcommand{\cm}{\mbox{\ensuremath{\mathrm{~cm}}}}

\newcommand{\km}{\mbox{\ensuremath{\mathrm{~km}}}}

\newcommand{\kpc}{\mbox{\ensuremath{\mathrm{~kpc}}}}
\newcommand{\Mpc}{\mbox{\ensuremath{\mathrm{~Mpc}}}}
\newcommand{\s}{\mbox{\ensuremath{\mathrm{~s}}}}
\newcommand{\ks}{\mbox{\ensuremath{\mathrm{~ks}}}}

\newcommand{\keV}{\mbox{\ensuremath{\mathrm{~keV}}}}
\newcommand{\erg}{\mbox{\ensuremath{\mathrm{~erg}}}}

\newcommand{\arcm}{\ensuremath{\mathrm{^\prime}}}
\newcommand{\arcs}{\arcm\hskip -0.1em\arcm}

\newcommand{\pcmsq}{\mbox{\ensuremath{\mathrm{~cm^{-2}}}}}
\newcommand{\pMpc}{\ensuremath{\mathrm{\Mpc^{-1}}}}
\newcommand{\ps}{\ensuremath{\mathrm{\s^{-1}}}}

\newcommand{\ergps}{\ensuremath{\mathrm{\erg \ps}}}
\newcommand{\flux}{\ensuremath{\mathrm{\erg \ps \pcmsq}}}

\newcommand{\kmpspMpc}{\ensuremath{\mathrm{\km \ps \pMpc\,}}}
\newcommand{\LT}{\mbox{\ensuremath{\mathrm{L-T}}}}
\newcommand{\ML}{\mbox{\ensuremath{\mathrm{M-L}}}}

\newcommand{\MnT}[1]{\mbox{\ensuremath{\mathrm{M_{#1}-T}}}}

\newcommand{\MT}{\mbox{\ensuremath{\mathrm{M-T}}}}
\newcommand{\MgT}{\mbox{\ensuremath{\mathrm{M_{g}-T}}}}

\newcommand{\LCDM}{$\Lambda$CDM~}

\begin{document}

%\special{!userdict begin /bop-hook{gsave 200 30 translate 65 rotate
% /Times-Roman findfont 216 scalefont setfont 0 0 moveto 0.93 setgray
% (DRAFT) show grestore}def end}

\title[Evolution of Cluster X-ray Scaling Relations]{The evolution of the cluster X-ray scaling relations in the WARPS sample at $0.6<z<1.0$.}
\author[B.J. Maughan \etal]
  {B. J. Maughan,$^{1,2}$\thanks{E-mail: bmaughan@cfa.harvard.edu}\thanks{Chandra Fellow}
    L. R. Jones,$^1$ H. Ebeling,$^3$ and C. Scharf $^4$\\ 
  $^1$School of Physics and Astronomy, The University of Birmingham,  Edgbaston, Birmingham B15 2TT, UK.\\
  $^2$Harvard-Smithsonian Center for Astrophysics, 60 Garden St, Cambridge, MA 02140, USA.\\
  $^3$Institute for Astronomy, 2680 Woodlawn Drive, Honolulu, HI 96822, USA.\\ 
  $^4$Columbia Astrophysics Laboratory, MC 5247, 550 West 120th St., New York, NY 10027, USA.\\
}

%\date{}
\maketitle

\begin{abstract}
The X-ray properties of a sample of 11 high-redshift ($0.6<z<1.0$) clusters
observed with \Chandra\ and/or \XMM\ are used to investigate the evolution
of the cluster scaling relations. The observed evolution in the
normalisation of the \LT, \MT, \MgT, and \ML\ relations are
consistent with simple self-similar predictions, in which the properties of
clusters reflect the properties of the universe at their redshift of
observation. Under the assumption that the model of self-similar evolution
is correct and that the local systems formed via a single spherical
collapse, the high-redshift \LT\ relation is consistent with the high-z
clusters having virialised at a significantly higher redshift than the local
systems. The data are also consistent with the more realistic scenario of
clusters forming via the continuous accretion of material. 

The slope of the \LT\ relation at high-redshift ($B=3.32\pm0.37$) is
consistent with the local relation, and significantly steeper then the
self-similar prediction of $B=2$. This suggests that the same
non-gravitational processes are responsible for steepening the local and
high-z relations, possibly occurring universally at $z\gta1$ or in the early
stages of the clusters' formation, prior to their observation. 

The properties of the intra-cluster medium at high-redshift are found to be
similar to those in the local universe. The mean surface-brightness profile
slope for the sample is $\beta=0.66\pm0.05$, the mean gas mass fractions
within $R_{2500(z)}$ and $R_{200(z)}$ are $0.069\pm0.012$ and $0.11\pm0.02$
respectively, and the mean metallicity of the sample is $0.28\pm0.11\Zsol$.

\end{abstract}


\section{Introduction} \label{c4sect_intro}
A simple and useful model to describe galaxy clusters is that they are
self-similar. In this model, clusters form via the collapse of the most
overdense regions in the early universe, and the cluster baryons are heated
only by gravitational processes (compression and shock heating) during the
collapse. The properties of clusters at high redshift are then identical to
those of their low-redshift counterparts, apart from scaling factors
reflecting the increase of the mean density of the universe with
redshift. This scaling with redshift has been termed weak self-similarity
\citep[\egc][]{bow97}. In the strongest form of the model, the mass
profiles of all clusters at the same epoch follow the same shape
independent of their total mass. The self-similar model then allows
properties of clusters of different masses and at different redshifts to be
related to one-another according to simple scaling laws. 

X-ray observations provide a powerful way of measuring cluster properties,
and have provided a wealth of evidence that galaxy clusters do not scale
self-similarly with mass (or, by proxy, temperature) in the local
universe. For example; the slope of the X-ray luminosity-temperature (\LT)
relation is steeper than self-similar predictions
\citep[\egc][]{mar98a,arn99}; the slopes of the gas-density and
surface-brightness profiles are shallower in cooler systems
\citep[\egc][]{llo00,san03}; and the entropy in cluster cores is higher
than predicted \citep[\egc][]{pon99,pon03}. These departures from
self-similarity are generally taken as evidence for the importance of
non-gravitational contributions (such as heating by AGN or radiative
cooling) to the energy budget of clusters.  

In this paper we address the open question of whether the simple evolution
of the cluster scaling relations predicted by the self-similar model is
obeyed. We define evolution as any change with respect to the local scaling
relations, and compare any such evolution with the self-similar
predictions. The results of early studies of the evolution of the \LT\
relation were consistent with little or no evolution
\citep[\egc][]{mus97b,don99,fai00}. More recent studies with \Chandra\ and
\XMM\ have found significant evolution in the \LT\ relation
\citep{vik02,ett04,lum04c}. This change is due to the availability of
larger samples of clusters at higher redshifts, and differences in the
assumed cosmological model. The measured evolution is larger in a \LCDM
cosmology than in the Einstein de-Sitter models which were assumed in many
of the earlier studies \citep[\egc][]{arn02a,lum04c}. 

Here we present an analysis of a sample of 11 clusters in the redshift
range  $0.6< z \le1.0$ drawn from the Wide Angle \ROSAT\ Pointed Survey
\citep[WARPS:][]{sch97,per02}. A statistically complete, flux-limited
($F_X(0.5-2\keV)>6.4\times10^-13\flux$)
sample of 13 WARPS clusters was originally observed with \Chandra\ and/or
\XMM, but two of the \XMM\ observations were rendered unusable due to
extremely high background levels, leaving the 11 clusters discussed
here. Of these clusters, two now fall below the flux limit due to point
source contamination in the original \ROSAT\ observations, however the set
of $11$ clusters used in this work should be fairly unbiased. These
clusters' properties, and the scaling relations derived from them, are
compared to those of other samples at high and low redshift, and with the
predictions of different cluster-formation models. 

A \LCDM cosmology of $H_0=70\kmpspMpc\equiv100h_{70}\kmpspMpc$, and
$\OM=0.3$ ($\Omega_\Lambda=0.7$) is adopted throughout, with the convention
that $\OM$ represents the present-day matter density, while $\Om(z)$
represents its value at redshift $z$. All errors are quoted at the $68\%$
level.

\section{Data Analysis}\label{c4sect_anal}
The standard data reduction steps were followed for both the \Chandra\ and
\XMM\ observations, and are discussed in detail in \citet{mau03a}
(\Chandra) and \citet{mau04a} (\XMM). Data reduction was performed with
Ciao 3.2.2 and SAS 6.1 for \Chandra\ and \XMM\ respectively. In summary,
the data were filtered to 
remove high-background periods, and a surface-brightness profile was
extracted (with point sources excluded) to determine the extent of the
cluster emission. The detection radius (\rd) of the cluster was then
defined as the radius outside which no further emission was detected at the
$3\sigma$ level in the surface-brightness profile. Spectra were extracted
from within the detection radius, and were fit in the $0.4-7\keV$ band with
an absorbed \MEKAL\ model \citep{kaa93}. The low energy cutoff was chosen
to minimise calibration uncertainties at low energies for both \XMM\ and
\Chandra. Ignoring energies above $7\keV$ has little effect on \Chandra\
data due to the low effective area at those energies, and avoids
instrumental fluorescent lines at $\sim8\keV$ which vary spatially across
the \XMM\ PN detector. The absorbing column was frozen at the Galactic
value determined from $21\cm$ radio observations \citep{dic90} during the
spectral fits. When the spectra were of sufficient quality, the absorbing
column was also fit, and found to be consistent with the Galactic
value. The surface-brightness distribution of each system was modeled with
a two-dimensional $\beta$-model \citep{cav76}.

The issue of background subtraction was carefully considered when modeling
both the spectra and the surface-brightness distributions. In all cases the
background was measured from the same observation in a region as close to
the source as possible, while avoiding contaminating source emission. Due
to the higher background levels in the \XMM\ observations, the background
in the \XMM\ surface-brightness models included two components to account
for the flat and vignetted background components
\citep{mau04a}. Consistency checks were performed for several clusters from
the sample using backgrounds derived from different regions of the source
datasets and from blank-sky datasets \citep[\egc][]{mau03a,mau04a}. The
derived properties were generally independent of the background used. Point
sources were masked out from all source and background regions during the
analyses. 

During the spectral-fitting process, the model redshifts were frozen at the
values derived from optical spectroscopy, and best-fitting temperatures
were found with the metallicity (Z) fixed at $0.3\Zsol$. The metallicity
parameter was then allowed to vary in addition to the model temperature and
normalisation, and its best-fitting value was found. In all cases, the
best-fitting model temperatures obtained with and without the metallicity
free to vary agreed within $1\sigma$. As the metallicities were not always
well constrained, the temperatures derived with metallicity fixed at
$0.3\Zsol$ are used throughout this work. The effective area of the
instruments was taken into account in the spectral modeling by using
ancillary response files (ARFs) generated for the cluster positions, and
weighted by the spatial distributions of source photons.

For the purposes of spectral fitting, additional filtering was applied to
the \XMM\ data retaining only events with FLAG and PATTERN parameters equal
to zero. These correspond to events detected in single pixels which were
not close to CCD gaps. The energy calibration of these events is the most
reliable. The loss of effective area to CCD gaps, bad pixels, and excluded
point sources within the spectral extraction region was accounted for by
correcting the normalisation of the ARFs. While the ARF files already
account for these losses in principal, they do not take into account the
surface brightness distribution of the source. We included this effect in
our correction as follows. For each cluster, a background-subtracted radial
profile of the source region was produced, excluding ``dead regions'' (CCD
gaps, bad pixels and point sources). In each radial bin, the measured flux
was used to predict the number of counts that would have been detected if
there were no dead regions. These counts were summed over the source
region, and the ARF normalisation was scaled by the ratio of the detected
counts to the predicted total counts if there were no dead regions. This
process was not required for the \Chandra\ observations as the spectral
extraction regions were unaffected by dead regions. The spectral
redistribution matrix files were generated with SAS version 6.1 using the
calibration database appropriate for that release, which included improved
calibration of the PN response at low energies. 

All of the clusters in our sample, with the exception of ClJ0046.3$+$8530, were the target of their observations and were located close to the optical axis. However ClJ0046.3$+$8530 was observed serendipitously near the edge of the field of view in two \XMM\ observations, which introduced some additional calibration issues due to the broader PSF. The analysis of this system is discussed in detail in \citet{mau04b}.

The slope ($\beta$) and core radius (\rc) of the gas-density profile were derived from a two-dimensional elliptical $\beta$-model fitted to an image of the cluster emission, including the effects of vignetting and the PSF. The ellipticity of the model was defined as $e=(1-b/a)$, where $a$ and $b$ are the semi-major and -minor axes respectively. With the central gas density derived from the normalisation of the best-fitting spectral model \citep[\egc][]{mau03a}, it was then possible to derive the gas-mass profile assuming spherical symmetry. The total mass profile was then derived under the assumptions of isothermality, hydrostatic equilibrium, and spherical symmetry. From this total mass profile, an overdensity profile (with respect to the critical density $\rho_c(z)$ at the cluster's redshift) was derived, enabling the measurement of overdensity radii, $R_\Delta$. Here $R_\Delta$ refers to the radius within which the mean density is $\Delta\rho_c(z)$, and $\Delta$ is an overdensity factor. We take $\Delta$ to be a function of redshift, as explained in \textsection \ref{c4sect_theory}, and denote radii defined in this way as \eg\ $R_{200(z)}$ where $\Delta=200$ at $z=0$.

\subsection{Computation of errors}
The uncertainties on the derived properties of a cluster were obtained in
the following way. Distributions of the derived properties were computed
from 10,000 randomisations of the observed properties within their
uncertainty distributions. The $1\sigma$ confidence limits on each of the
derived properties were then obtained from the $\pm34$ percentiles about
the best-fit value. This method treats the statistical uncertainties in
extrapolating properties to different radii self-consistently, but does not
account for any systematic uncertainties in extrapolating properties beyond
the limits of the data. A significant source of systematic uncertainty is
the assumption of isothermality. Data of sufficient quality to measure
temperature profiles were only available for two clusters in the sample,
and in those cases the assumption of isothermality was justified. The
systematic effect that undetected temperature gradients could have on
derived masses is discussed in \textsection \ref{c4sect_MT} and
\textsection \ref{c4sect_MgT}.

An additional consideration is the well known positive correlation between the $\beta$ and \rc\ surface-brightness profile parameters. Models with large $\beta$ and \rc\ are similar to those with small values of the parameters. The error treatment described above assumes that all errors are independent. The effect of this assumption was investigated in the following way: A simple cluster image was simulated by adding Poisson noise to a two-dimensional $\beta$-profile image. A $\beta$-model was then fit to this image, and a two-dimensional probability distribution of $\beta$ and \rc\ values was generated. Cluster properties were then derived for an assumed temperature and \MEKAL\ normalisation, and the uncertainties were computed by sampling pairs of  $\beta$ and \rc\ values from the two-dimensional probability distribution, thereby accurately reflecting their correlated errors. The uncertainties on all other parameters were assumed to be negligible. This process was repeated assuming uncorrelated  $\beta$ and \rc\ errors, and also assuming negligible errors on \rc. It was found that the latter two methods gave similar uncertainties on cluster properties, and these uncertainties were approximately twice as large as those derived using the true, correlated errors on $\beta$ and \rc. In fact, the uncertainties on the properties of the observed clusters are dominated by the measurement errors on the temperature, and the choice between these methods makes a negligible contribution to the error budget. With this in mind, the measurement errors on \rc\ were ignored in the calculation of the errors on all cluster properties.

\begin{figure*}
\begin{center}
\scalebox{0.6}{\includegraphics*[angle=270]{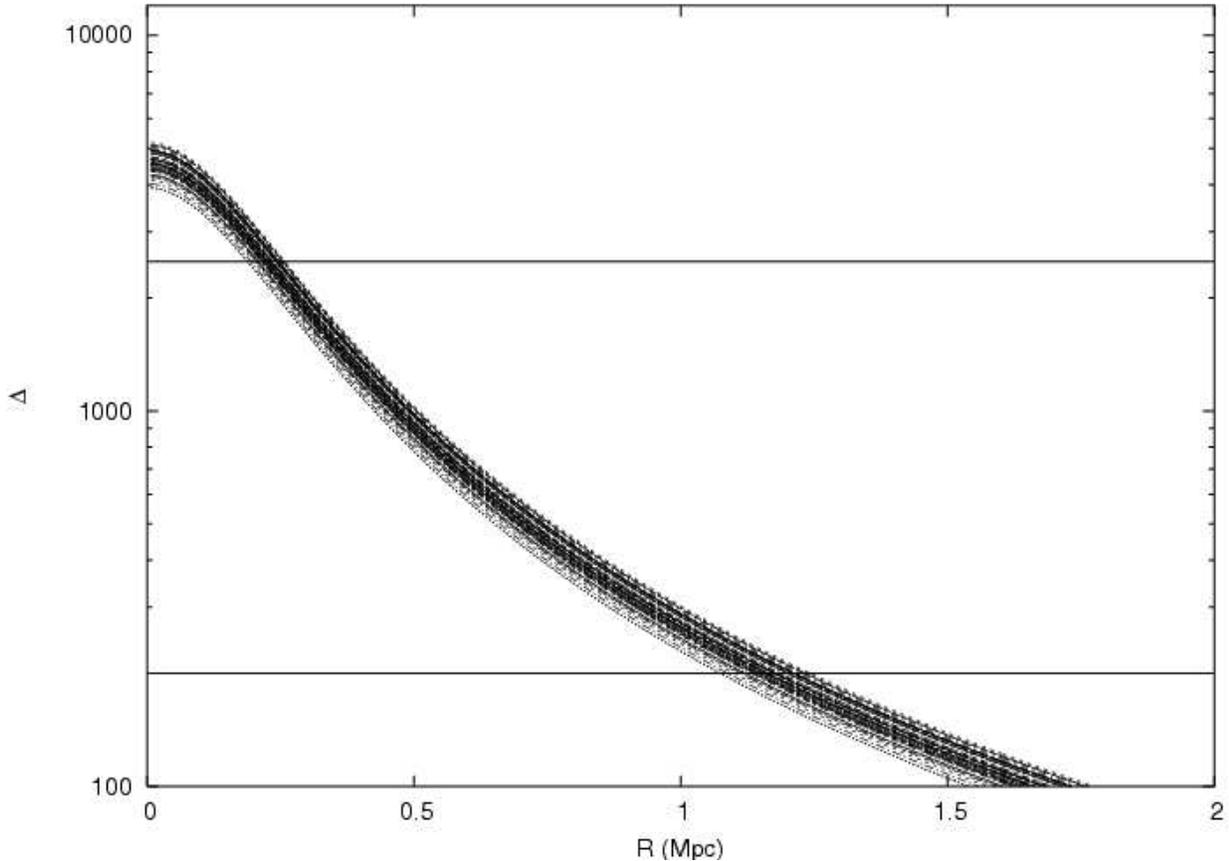}} \\
\caption[50 sample overdensity profiles generated for a distribution of core radii.]
{\label{c4fig_densprof}50 sample overdensity profiles generated for a distribution of $\beta$ values. The horizontal lines mark overdensities of 200 and 2500. The overdensity at R is defined as the ratio of the mean density within R to the critical density.}
\end{center}
\end{figure*}

An interesting consequence of this self-consistent treatment of errors is that the fractional uncertainty on $R_{2500}$ is in general significantly larger than that on $R_{200}$. This is in spite of the fact that the data for all of the clusters extend beyond $R_{2500}$, but must be extrapolated to $R_{200}$ (see Table \ref{c4tab_prop}). The reason for this is that without including any systematic uncertainties due to the extrapolation, the distribution of $\beta$ values has a larger effect on the shape of the density profile at small radii than large ($>>\rc$) radii. Fig. \ref{c4fig_densprof} illustrates this point, showing a range of 50 overdensity profiles derived for a $5.5\keV$ cluster at $z=0.833$ with $\rc=250\kpc$ and values of $\beta$ randomly drawn from the distribution $0.67\pm0.07$. These uncertainties on $\beta$ are the average fractional uncertainties on our sample, and so reflect the typical uncertainty in the shape of the overdensity profiles. In these simulations, the fractional error on $R_{2500}$ is $0.11$, while that on $R_{200}$ is $0.05$. The self-consistent treatment of the uncertainties can thus result in larger fractional error on quantities derived within the extent of the data than those extrapolated to large radii beyond the data. While the measurement errors on $kT$ have a large effect on the uncertainty in the normalisation of the overdensity profiles, this source of uncertainty is not important for this comparison as it does not affect the shape of the profiles.

\section{The Sample} \label{c4sect_samp}
\begin{figure*}
\begin{center}
\includegraphics[width=8.5cm]{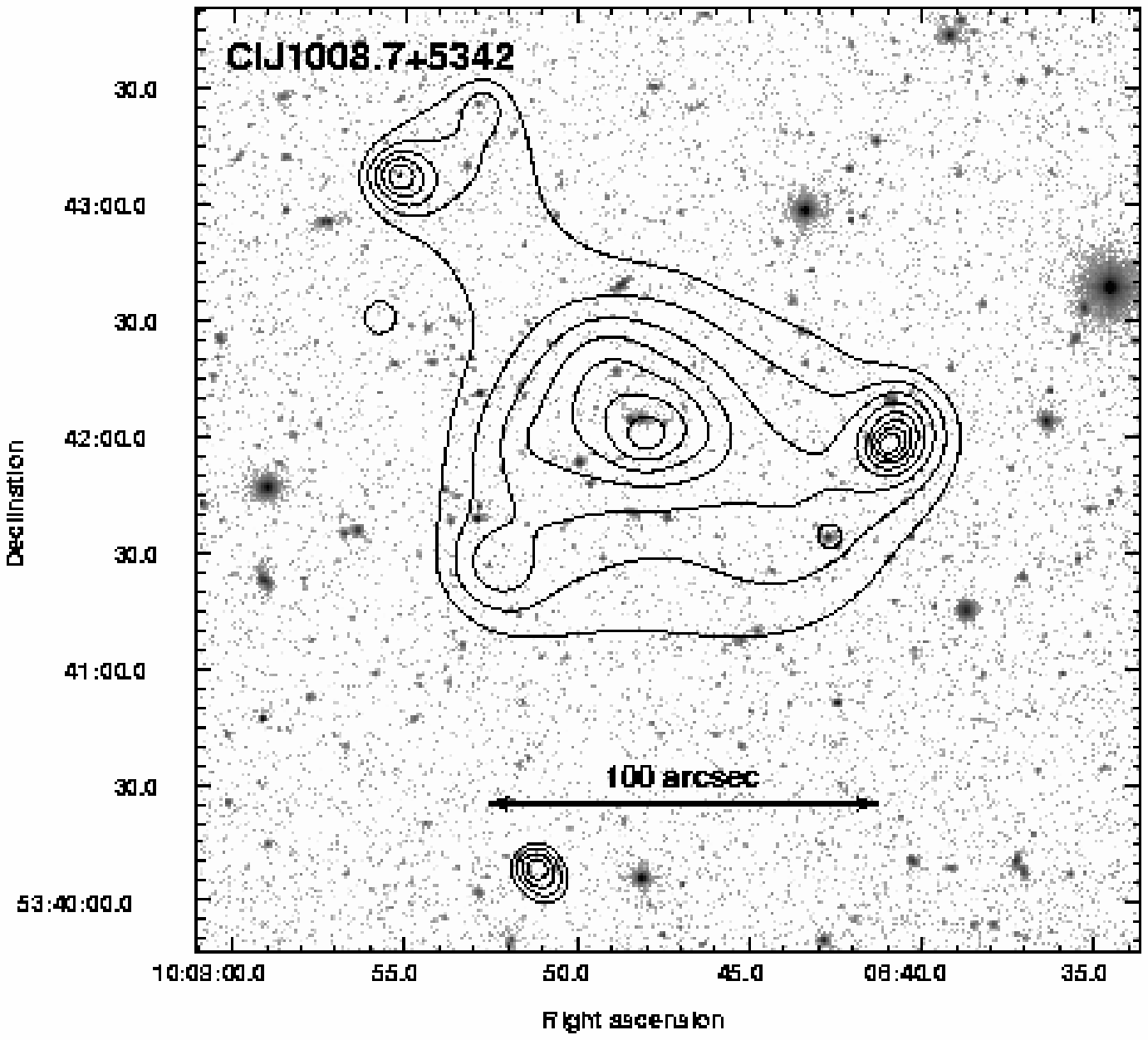}
\includegraphics[width=8.5cm]{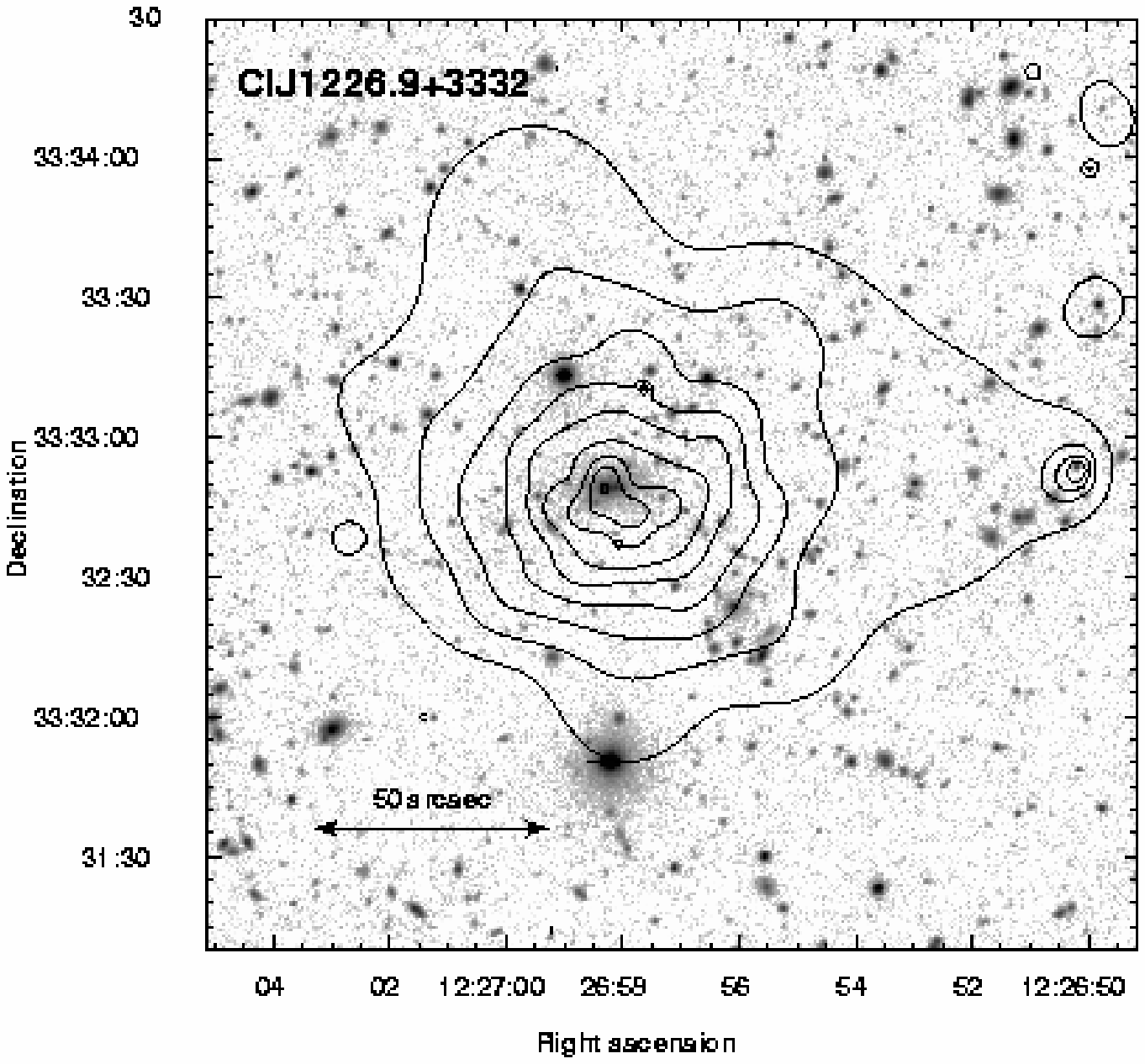}\\
\includegraphics[width=8.5cm]{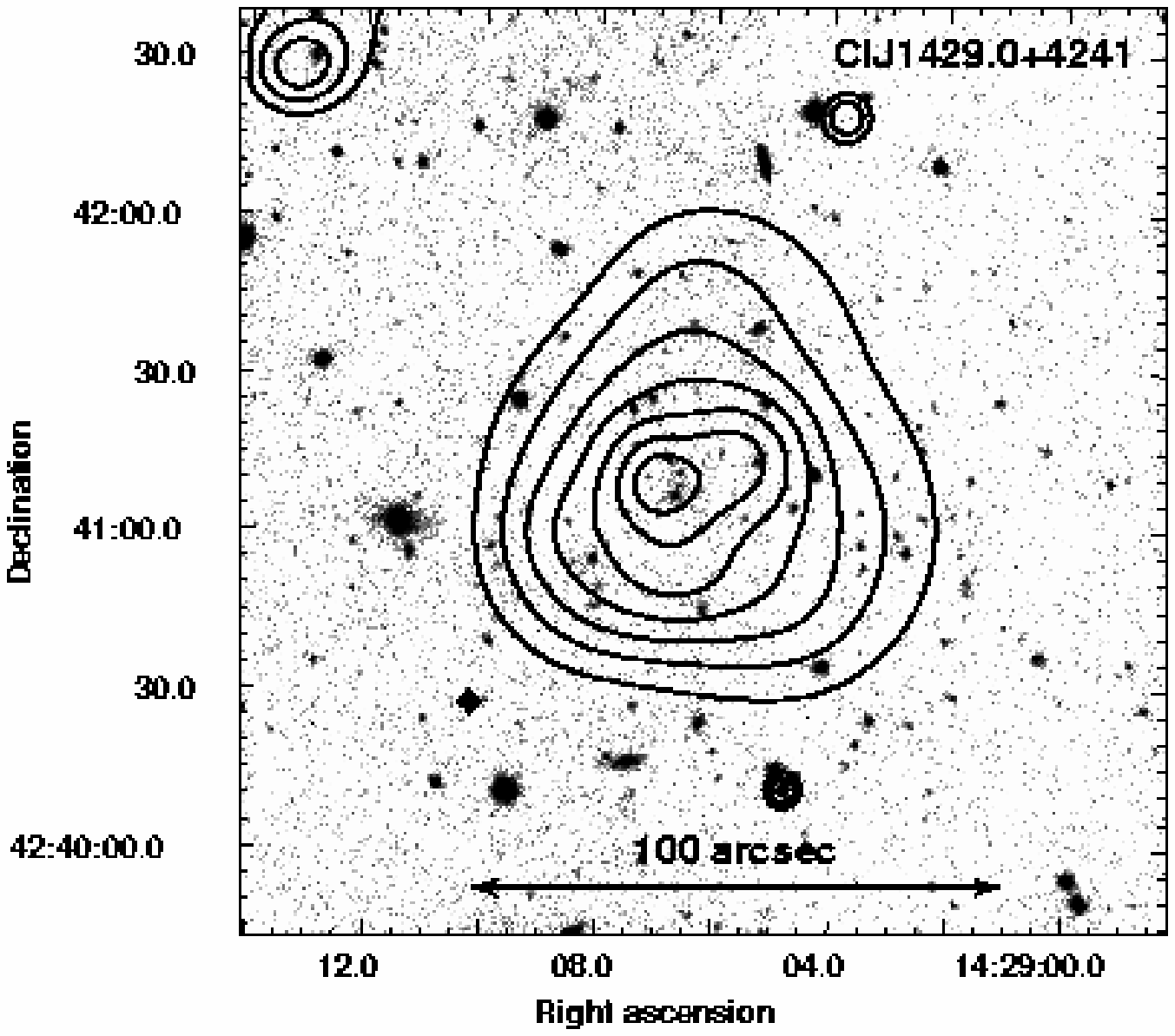}
\includegraphics[width=8.5cm]{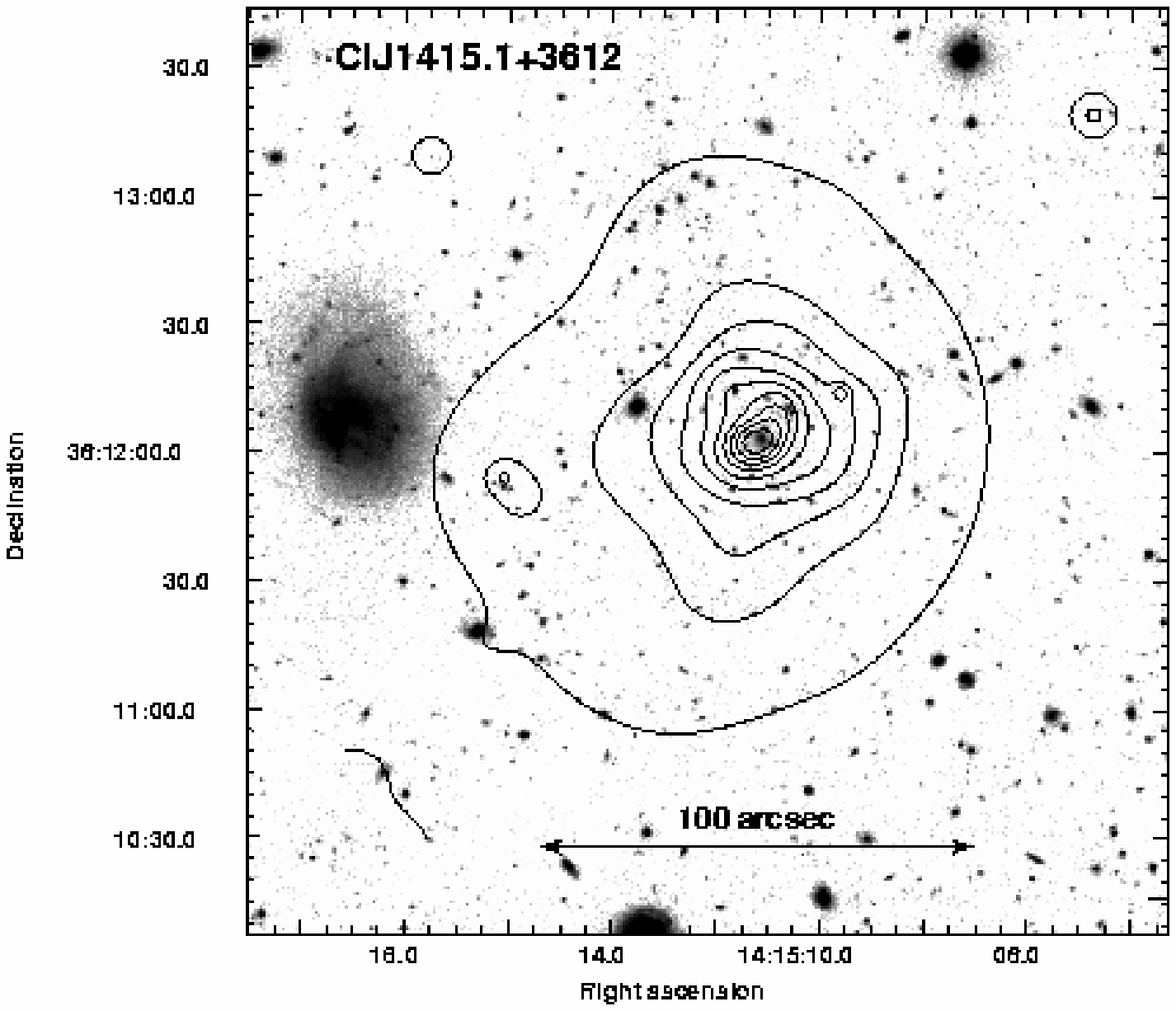}\\
\caption[Contours of X-ray emission overlaid on optical images of the WARPS high-redshift sample.]{\label{c4fig_overlays}Contours of X-ray emission overlaid on optical images of the WARPS high-redshift sample. The pass-bands and telescopes used to produce the optical images varied, with the I, R or Z band, and  Keck-II, the 4.2m William-Herschel Telescope, Subaru or the University of Hawaii's 2.2m telescope used. Contours are taken from images that were adaptively smoothed (using the {\it asmooth} algorithm of \citet{ebe05}) such that all features are significant at the $99\%$ level, and are logarithmically spaced, with the lowest contour a factor of $1.5$ above the background. The X-ray observatory used, and other information about the clusters is given in Table \ref{c4tab_prop}.}
\end{center}
\end{figure*}

In this section each cluster in the sample is discussed briefly, highlighting any unusual or interesting aspects of its analysis and properties. Images of each of the clusters with contours of the adaptively smoothed X-ray emission overlaid are shown in Fig. \ref{c4fig_overlays} and the observed properties of the sample are summarised in Table \ref{c4tab_prop} in redshift order. The detection radii of each cluster as a fraction of $R_{2500(z)}$ and $R_{200(z)}$ are also given in Table \ref{c4tab_prop}. The method used to define and measure these overdensity radii is described in \textsection \ref{c4sect_theory} and their values are given in Table \ref{c4tab_summary} for each cluster.

\begin{figure*}
\begin{center}
\includegraphics[width=8.5cm]{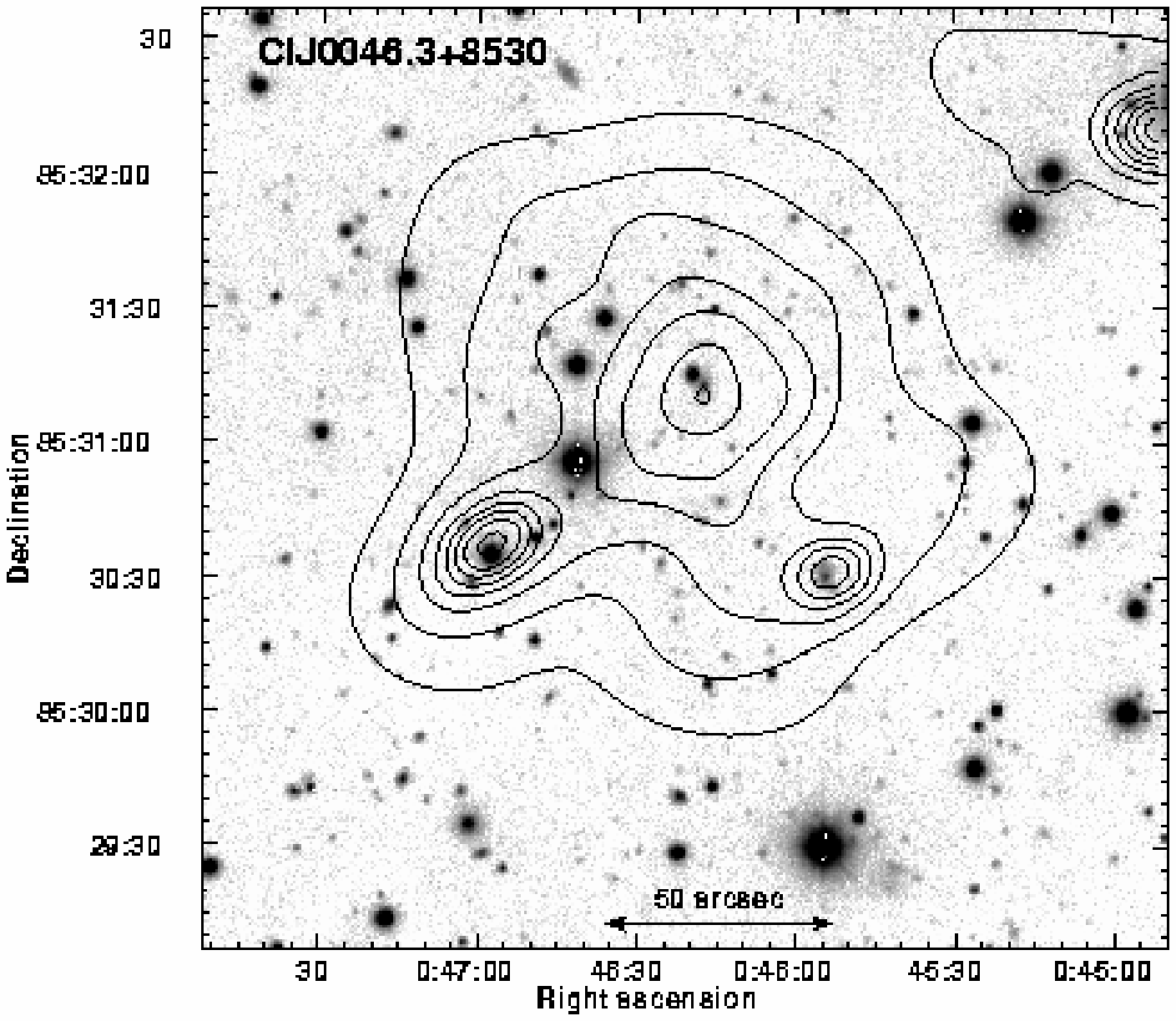}
\includegraphics[width=8.5cm]{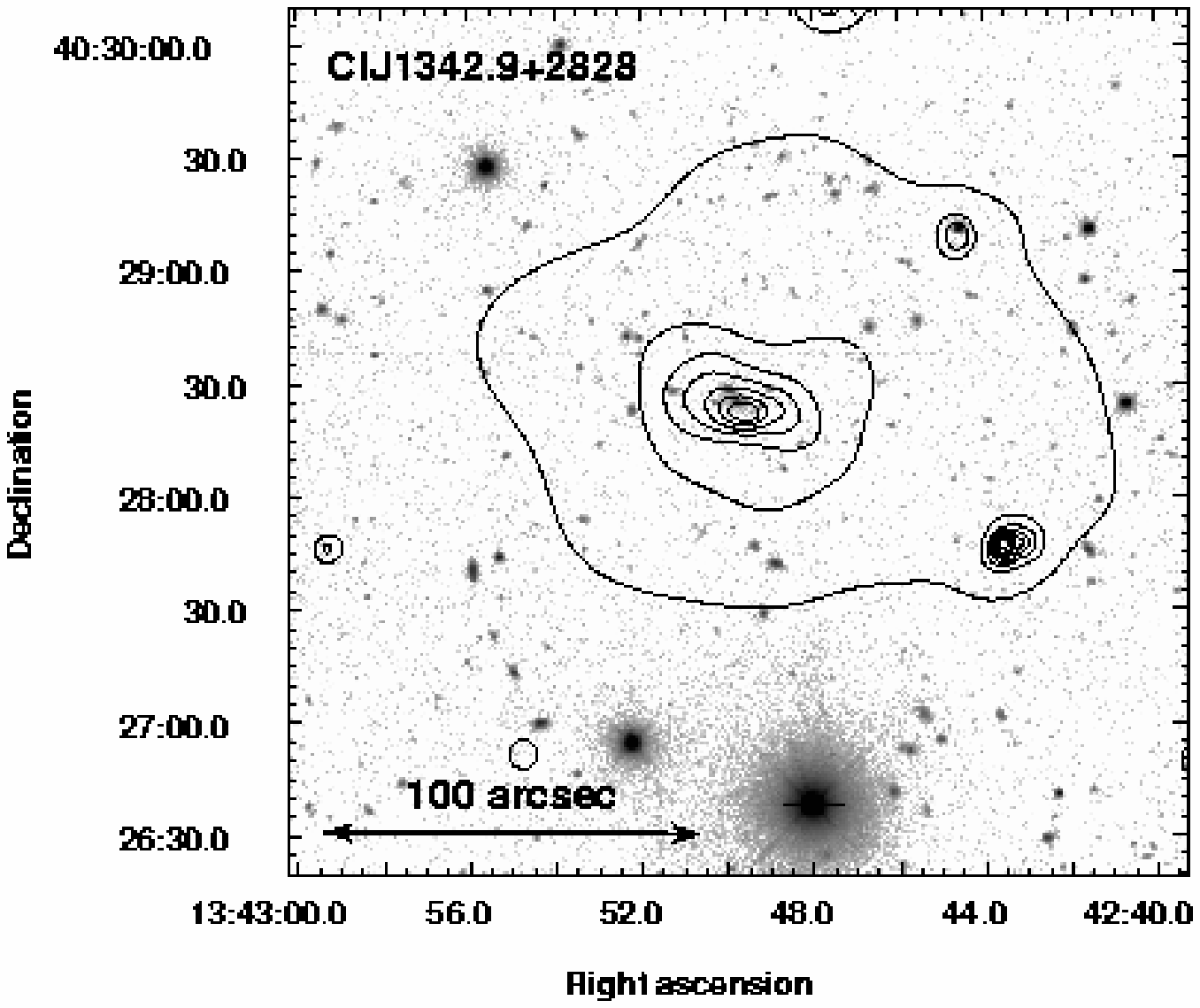}\\
\includegraphics[width=8.5cm]{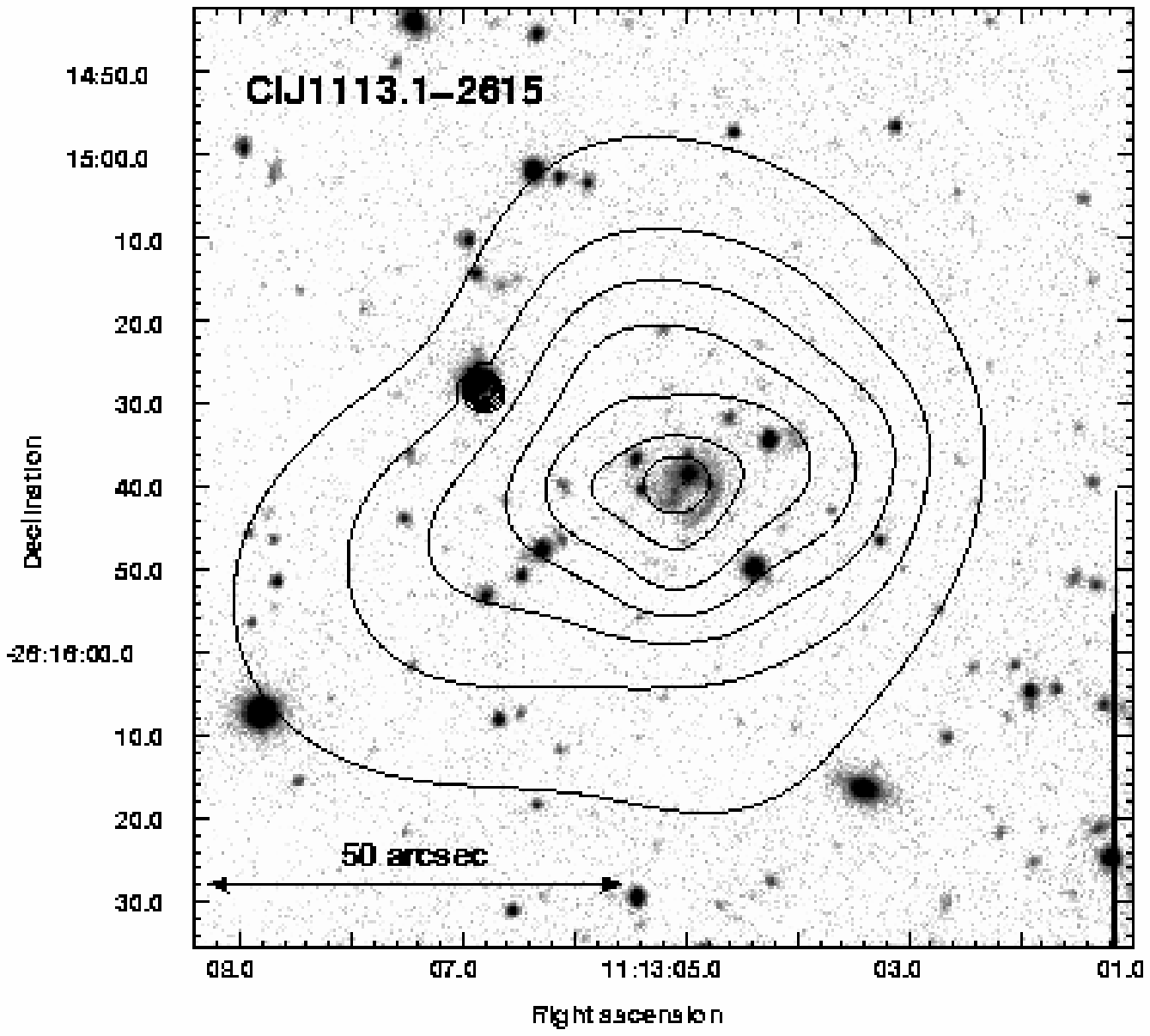}
\includegraphics[width=8.5cm]{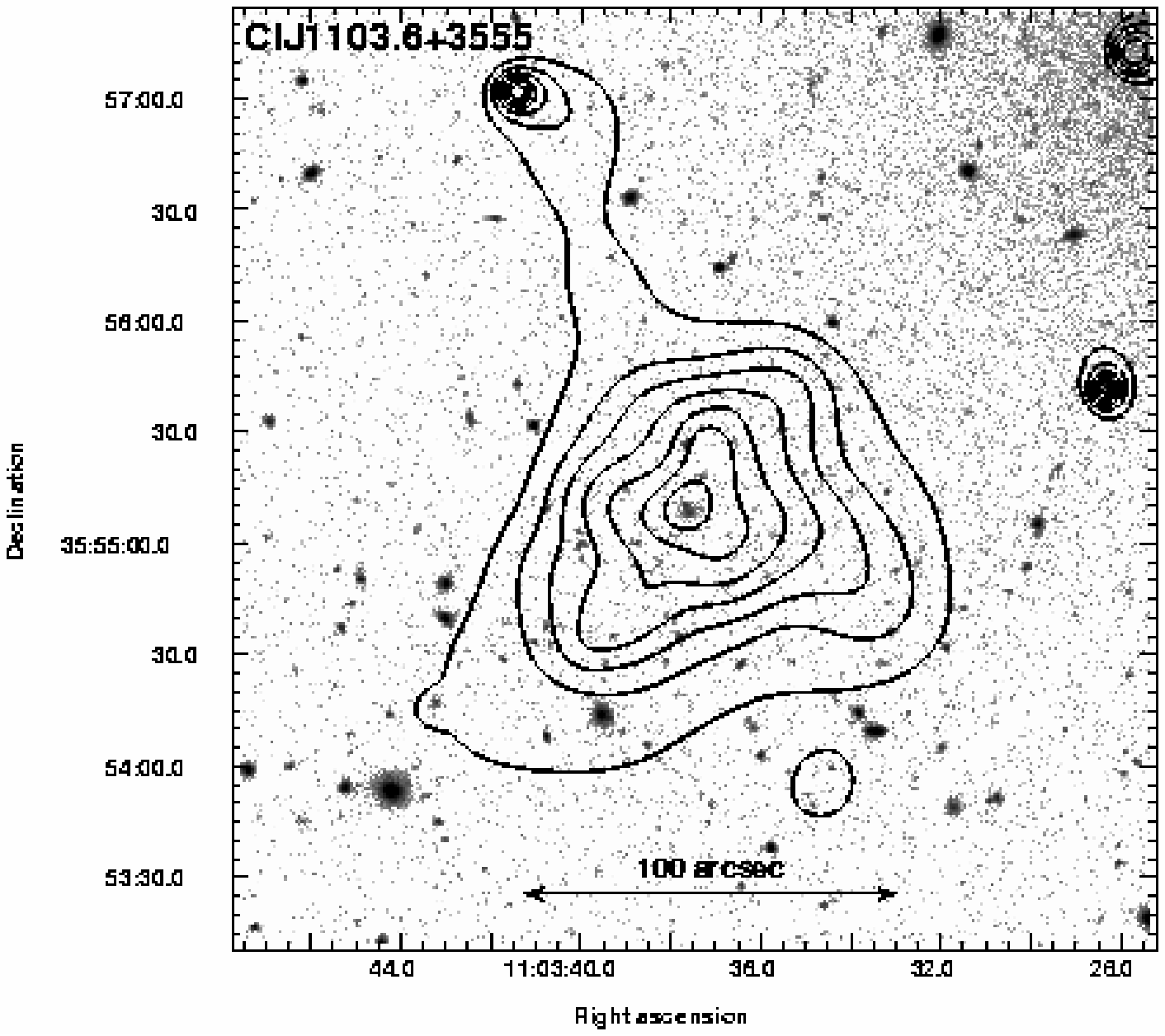}\\
\includegraphics[width=8.5cm]{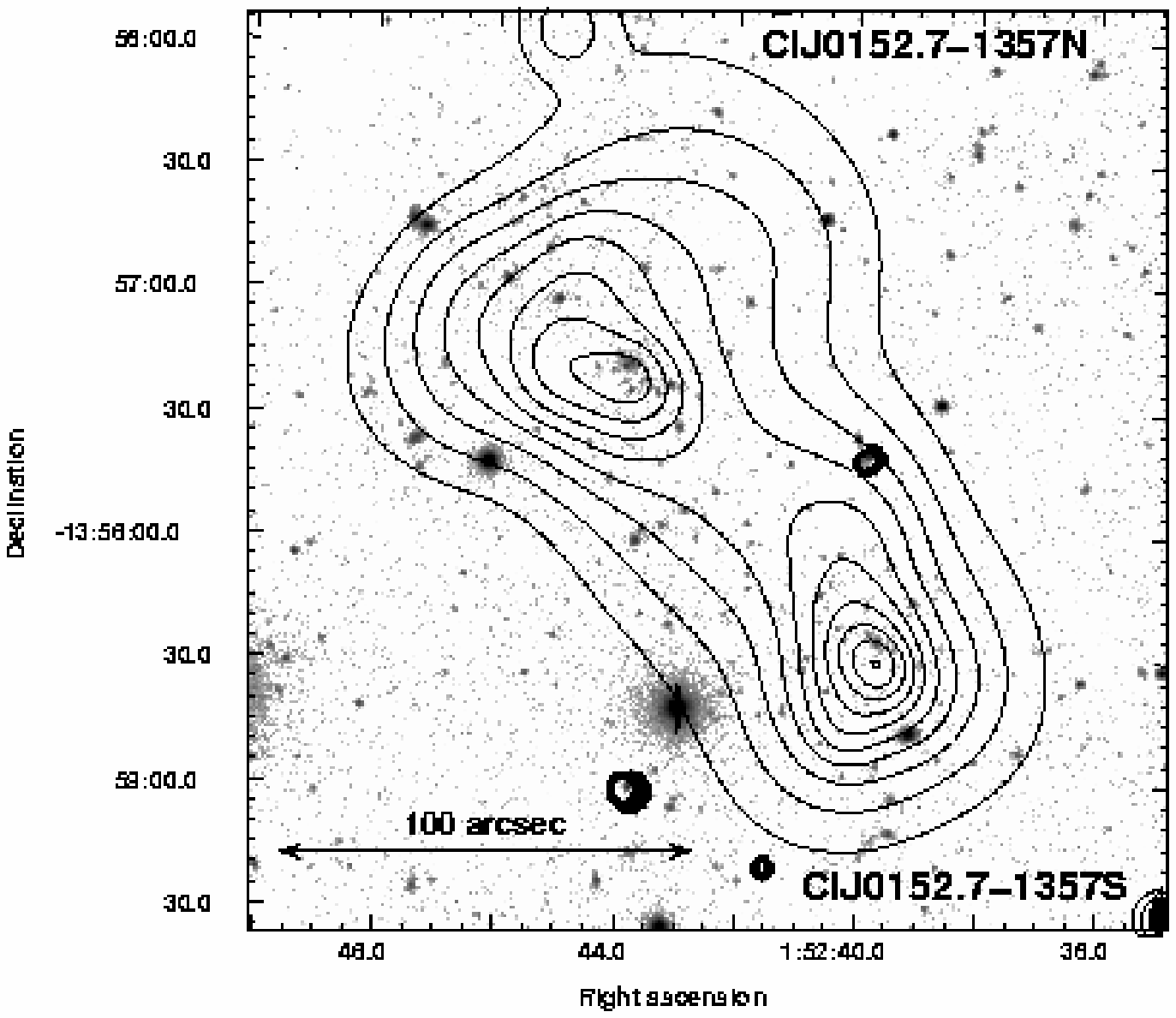}
\includegraphics[width=8.5cm]{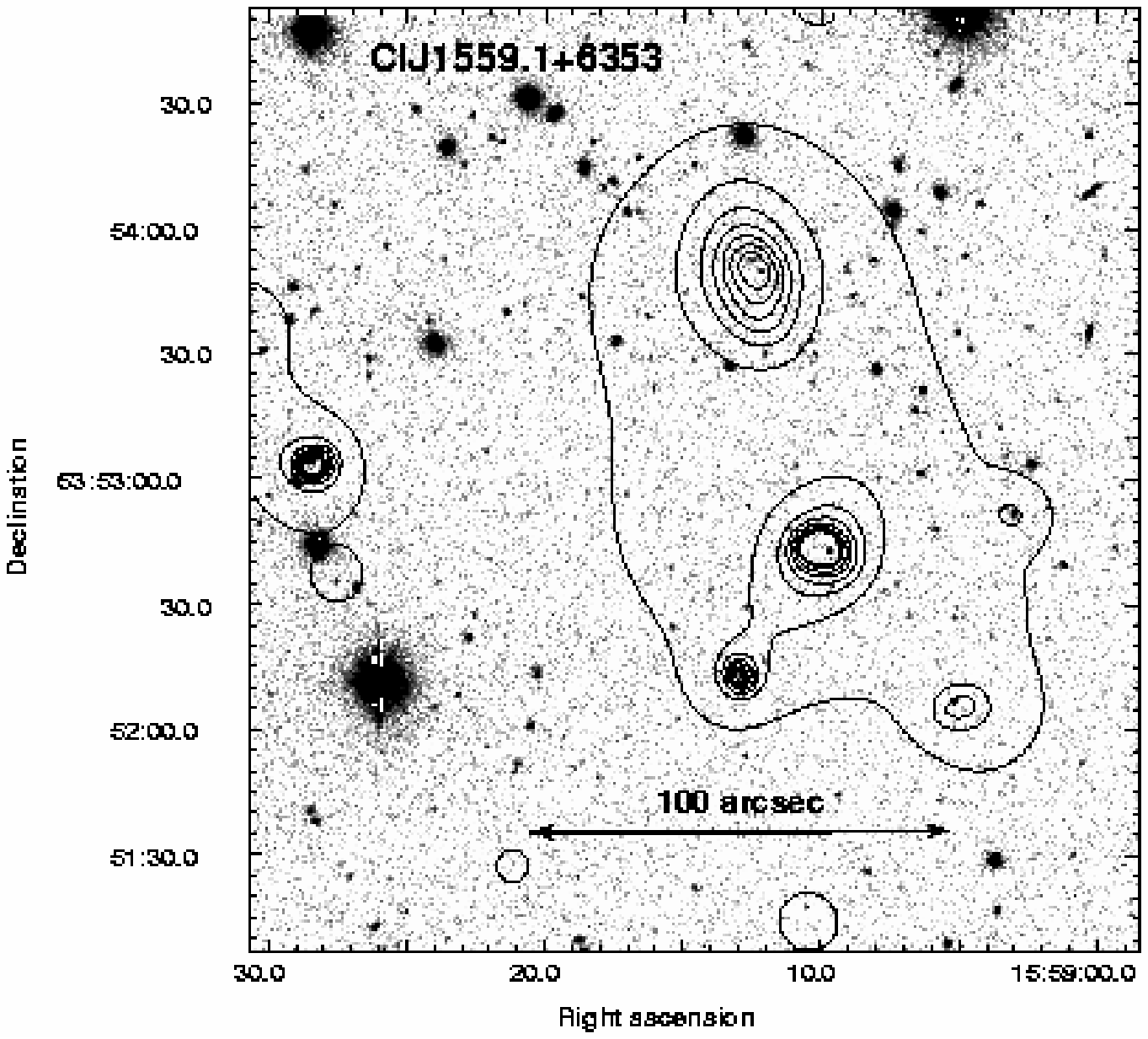}\\
\end{center}
\end{figure*}

\noindent{\bf ClJ0046.3$+$8530.} 
This cluster was observed serendipitously $11\arcm$ off axis in two consecutive \XMM\ observations of the open star cluster NGC 188. These data are discussed in detail in \citet{mau04b}. The cluster has a reasonably relaxed morphology, and its temperature profile and hardness-ratio map are consistent with isothermality out to $70\%$ of the virial radius, within the statistical limits of the data.

\noindent{\bf ClJ1342.9$+$2828.}
The \XMM\ observation of this system shows a core region that is elongated in the East-West direction, with two possible X-ray peaks. This morphology is indicative of a late-stage merger.

\noindent{\bf ClJ1113.1$-$2615.}
The \Chandra\ observation of this cluster, which is discussed in detail in \citet{mau03a}, shows a reasonably relaxed, although slightly elliptical, morphology.

\noindent{\bf ClJ1103.6$+$3555.}
Observed with \XMM, this cluster has a fairly disturbed X-ray morphology, suggesting that it may not have dynamically relaxed after a recent merger. It also appears to be surrounded by more extended low surface-brightness emission.

\noindent{\bf ClJ0152.7$-$1357.}
This spectacular system is probably an early-stage merger between two equally massive clusters ClJ0152.7$-$1357N and ClJ0152.7$-$1357S and has been studied in some detail \citep{ebe00a,mau03a,huo04,jee05}. The \Chandra\ observation used here suggests that both of the clusters are reasonably relaxed, with ClJ0152.7$-$1357N the more elliptical. A recent, deep \XMM\ observation has detected some substructure in ClJ0152.7$-$1357N, while ClJ0152.7$-$1357S still appears relaxed (Maughan et. al. in preparation).

\noindent{\bf ClJ1559.1$+$6353.}
This cluster was observed by \XMM. The data show an elliptical morphology ($e=0.35$; see Table \ref{c4tab_prop}), suggesting that the system may not yet have fully relaxed back into hydrostatic equilibrium after its last merger event. A bright, variable point-source $\approx1\arcm$ South of the cluster centroid led to an overestimate of its \ROSAT\ flux in the WARPS (Horner et. al. in preparation). For this system, a very high, poorly constrained metallicity was measured, with relatively poor constraints on temperature (see Table \ref{c4tab_prop}). This is likely to be due to some remaining contamination from the bright source caused by the large wings of the \XMM\ point spread function, and to the fact that the spectrum had a lower signal-to-noise than most of the other systems discussed here.

\noindent{\bf ClJ1008.7$+$5342.}
The morphology of this cluster appears fairly relaxed in the \XMM\ observation. However, the ellipticity of the best-fitting surface-brightness model is $0.27$, indicating that the system may be disturbed to some degree.

\noindent{\bf ClJ1226.9$+$3332.}
This system has been the target of both \Chandra\ \citep{cag01} and \XMM\ \citep{mau04a} observations, and was found to be extremely hot, with a regular, relaxed morphology. Based on a $16\ks$ \XMM\ observation \citet{mau04a} measured a temperature profile and hardness-ratio map which were consistent with the cluster being isothermal out to $45\%$ of the virial radius. In addition, the temperature and luminosity measured from the \Chandra\ and \XMM\ data were found to be in good agreement \citep{mau04a}. The properties derived from the \XMM\ data are used throughout this work. A forthcoming deep \XMM\ observation will allow this massive cluster to be studied in unprecedented detail.

\noindent{\bf ClJ1429.0$+$4241.}
This cluster was observed serendipitously during an \XMM\ observation of the BL Lac H1426$+$428. Only the data from the MOS2 detector were useful for its study because the other detectors were in fast timing mode. The cluster's morphology is fairly disturbed, particularly in the centre, with a possible second X-ray peak to the West. During the two-dimensional surface-brightness modeling of this system, it was not possible to constrain the $\beta$ parameter. This is likely to be due to a combination of the system's morphology, its compactness compared with the \XMM\ PSF, and the relatively shallow imaging with the single MOS detector. The fit was thus performed with $\beta$ fixed at the canonical value of $0.67$, and the errors quoted on $\beta$ hereafter are the mean fractional errors of the rest of the sample ($9\%$).

\noindent{\bf ClJ1415.1$+$3612.}
At $z=1.03$, the most distant cluster in the sample. The \XMM\ data show the morphology to be relaxed. A deep ($78\ks$) \Chandra\ observation (Ebeling et. al. in preparation) of this system confirms the relaxed morphology, and rules out significant unresolved point source contamination in the \XMM\ data used here.

The measurement of reliable masses for these clusters based on the X-ray
data requires that they be in hydrostatic
equilibrium. Five systems (ClJ1342.9$+$2828, ClJ1103.6$+$3555,
ClJ0152.7$-$1357N, ClJ1559.1$+$6353 and ClJ1429.0$+$4241) are possibly
unrelaxed showing disturbed morphologies, possible substructure and/or
large ellipticities. The mass estimates for these systems are thus likely
to be less reliable than those for the more relaxed systems.

%\begin{sidewaystable*}
\begin{table*}
\scalebox{0.9}{
\begin{tabular}{ccccccccccccc} \hline 
Cluster & z & satellite & exposure & & \rd$^\dag$ & & scale & $kT$ & Z & $\rc$ & $\beta$ & $e$\\
\cline{5-7}
&  &  & (ks) & ($\arcs$) & ($R_{2500(z)}$) & ($R_{200(z)}$) & (kpc$/\arcs$) & (keV) & ($\Zsol$) & (kpc) & & \\ \hline

ClJ0046.3$+$8530$^a$ & $0.62$ & X & 44 & 88 & 2.5 & 0.63 & 6.81 & $4.4^{+0.5}_{-0.4}$ & $0.61^{+0.22}_{-0.19}$ & $137^{+30}_{-25}$ & $0.60^{+0.08}_{-0.03}$ & 0.07 \\

ClJ1342.9$+$2828 & $0.71$ & X & 33 & 109 & 4.3 & 0.89 & 7.19 & $3.7^{+0.5}_{-0.4}$ & $0.19^{+0.22}_{-0.19}$ & $172^{+28}_{-24}$ & $0.70^{+0.06}_{-0.05}$ & 0.34 \\

ClJ1113.1$-$2615$^b$ & $0.73$ & C & 65 & 50 & 1.4 & 0.37 & 7.24 & $4.7^{+0.9}_{-0.7}$ & $0.71^{+0.40}_{-0.31}$ & $106^{+9}_{-16}$ & $0.67^{+0.03}_{-0.05}$ & 0.20 \\

ClJ1103.6$+$3555 & $0.78$ & X & 36 & 95 & 2.9 & 0.72 & 7.43 & $6.0^{+0.9}_{-0.7}$ & $0.42^{+0.27}_{-0.21}$ & $141^{+21}_{-16}$ & $0.58\pm0.03$ & 0.18 \\

ClJ0152.7$-$1357N$^b$ & $0.83$ & C & 31 & 49 & 2.5 & 0.37 & 7.61 & $5.6^{+1.0}_{-0.8}$ & $0.33^{+0.27}_{-0.23}$ & $249^{+63}_{-33}$ & $0.73^{+0.13}_{-0.06}$ & 0.08 \\

ClJ0152.7$-$1357S$^b$ & $0.83$ & C & 31 & 37 & 1.3 & 0.31 & 7.61 & $4.8^{+1.1}_{-1.0}$ & $0.19^{+0.52}_{-0.19}$ & $123^{+28}_{-20}$ & $0.66^{+0.08}_{-0.06}$ & 0.00 \\

ClJ1559.1$+$6353 & $0.85$ & X & 19 & 73 & 2.6 & 0.72 & 7.54 & $4.1^{+1.4}_{-1.0}$ & $1.30^{+2.90}_{-0.70}$ & $67^{+34}_{-25}$ & $0.59^{+0.06}_{-0.11}$ & 0.35 \\

ClJ1008.7$+$5342 & $0.87$ & X & 14 & 98 & 5.4 & 1.00 & 7.72 & $3.6^{+0.8}_{-0.6}$ & $0.11^{+0.43}_{-0.11}$ & $170^{+47}_{-39}$ & $0.68^{+0.10}_{-0.08}$ & 0.27 \\

ClJ1226.9$+$3332$^c$ & $0.89$ & X & 17 & 100 & 2.2 & 0.60 & 7.76 & $10.6^{+1.1}_{-1.1}$ & $0.49^{+0.17}_{-0.17}$ & $113^{+9}_{-6}$ & $0.66^{+0.02}_{-0.02}$ & 0.14 \\

ClJ1429.0$+$4241 & $0.92$ & X & 44 & 50 & 1.5 & 0.40 & 7.84 & $6.2^{+1.5}_{-1.0}$ & $0.49^{+0.55}_{-0.44}$ & $97\pm9$ & $0.67\pm0.06^\ddag$ & 0.16 \\

ClJ1415.1$+$3612 & $1.03$ & X & 17 & 60 & 2.1 & 0.55 & 8.06 & $5.7^{+1.2}_{-0.7}$ & $0.45^{+0.39}_{-0.33}$ & $94^{+19}_{-14}$ & $0.67^{+0.06}_{-0.04}$ & 0.16 \\ \hline

\end{tabular}
}
\caption[Summary of the observed properties of the WARPS high-redshift sample.]{\label{c4tab_prop}Summary of the observed properties of the WARPS high-redshift sample. Column 3 indicates whether the observation used here was made with \Chandra\ (C) or \XMM\ (X). Column 4 gives the on-axis exposure time remaining after removal of high-background periods (the mean of the MOS and PN times is given for \XMM\ observations). Detailed analyses of several of the clusters can be found in $^a$\citet{mau04b}, $^b$\citet{mau03a} and $^c$\citet{mau04a}. $^\dag$The detection radius is given in units of $R_{2500(z)}$ and $R_{200(z)}$ for each cluster. These radii are defined in \textsection \ref{c4sect_theory} and given in Table \ref{c4tab_summary}. $^\ddag$The value of $\beta$ for this cluster was fixed at $0.67$ during the surface-brightness fitting, and the error quoted here is the mean fractional error on $\beta$ in the rest of the sample.}
%\end{sidewaystable*}
\end{table*}

\subsection{Comparison of \Chandra\ and \XMM\ Temperatures}
\begin{figure}
\begin{center}
\scalebox{0.45}{\includegraphics*[angle=270]{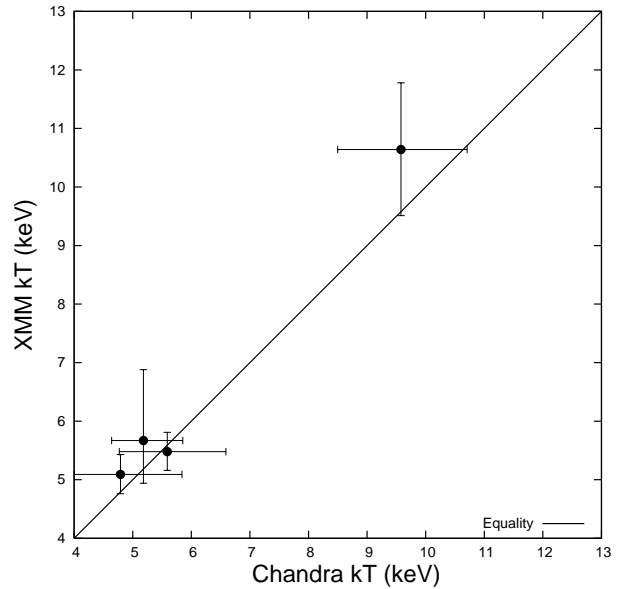}} \\
\caption[]{\label{c4fig_ktkt}Comparison of temperatures measured with \Chandra\ and \XMM\ for the four clusters in our sample observed by both.}
\end{center}
\end{figure}

In compiling a sample which includes observations performed with two different instruments, the accuracy of their cross-calibration is an important consideration. In particular, for the study of scaling relations, it is important that no systematic bias in the measured temperature is present due to the imperfect calibration of either instrument. For 4 of the clusters in the sample (ClJ0152.7$-$1357N, ClJ0152.7$-$1357S, ClJ1226.9$+$3332 and  ClJ1415.1$+$3612) observation made with both \Chandra\ and \XMM\ were available to us. This enabled us to measure temperatures in a consistent way with both observatories. The measured temperatures are plotted in  Fig. \ref{c4fig_ktkt}.

While limited by the small number of comparison points and the size of the
statistical uncertainties on the temperatures, Fig. \ref{c4fig_ktkt}
suggests no gross systematic disagreement between the \Chandra\ and \XMM\
temperatures. This evidence, along with the good agreement between the
WARPS \LT\ relation and the \Chandra\ \LT\ relation of \citet{vik02}
(discussed in detail in \textsection \ref{sect_v02}) indicate that the
cross calibration of the two satellites is not a significant problem in this
work. In studies of nearby clusters with high quality observations, some
systematic temperature disagreements between \Chandra\ and \XMM have been
found \citep[\egc][]{san05,vik05}. Given the evidence above, it appears
that any such effect is small compared to the statistical uncertainties in
this work.

\section{Cluster Scaling Theory} \label{c4sect_theory}
When comparing integrated cluster properties such as luminosity and mass with theoretical predictions, simulations, or other work, the radius within which the properties are measured is of great importance. It is common to use a fixed physical size as an outer radius, which has obvious benefits in terms of simplicity. A more sophisticated method is to define a radius with some knowledge of the cluster's mass profile so that the mean enclosed density is a fixed factor above the critical density of the universe. This method is more appropriate for comparisons with theoretical predictions, which predict the outer boundary of the virialised part of clusters in terms of a density contrast ($\Dv$). For instance, in an Einstein-de Sitter universe, the mean density of virialised systems is $\Dv=18\pi^2\rho_c$ at all redshifts \citep[\egc][]{bry98}. However, the value of $\Dv$ and its variation with redshift are cosmology dependent (as, of course, is $\rho_c(z)$). \citet{bry98} calculate the redshift dependence of $\Dv$ in a \LCDM cosmology for clusters which have just virialised, fitting the solution with the expression
\begin{eqnarray}\label{c4eqn_delta}
\Dv(z) & = & 18\pi^2 + 82(\Om(z)-1) - 39(\Om(z)-1)^2,
\end{eqnarray}
where $\Om(z)=\OM(1+z)^3/E(z)^2$, and $E(z)$ describes the redshift evolution of the Hubble parameter, given by
\begin{eqnarray}\label{c4eqn_ez}
E^2(z) & = & \OM(1+z)^3 + (1-\OM-\Omega_\Lambda)(1+z)^2 + \Omega_\Lambda.
\end{eqnarray}
Equation \ref{c4eqn_delta} is accurate to within $1\%$ in the range $0.1\le\Omega(z)\le1$ \citep{bry98}, which corresponds to all $z$ for $\OM=0.3$. 

The ideal method to follow would therefore be to measure cluster properties within radii corresponding to $\Dv(z)$ at all redshifts. This is not possible because X-ray measurements typically only extend to a fraction of the radius corresponding to $\Dv(z)$, even at low redshifts, requiring the extrapolation of measured properties by large factors. An alternative to this extrapolation is to work at a smaller radius, corresponding to a higher density contrast, $\Delta(0)$, at $z=0$. In the self-similar model, this contrast will scale with redshift according to 
\begin{eqnarray}\label{c4eqn_da}
\Delta(z)=\Delta(0)\frac{\Dv(z)}{\Dv(0)}.
\end{eqnarray}

In our analysis, all cluster properties are extrapolated to several different radii to enable comparison with other work. We note that for comparison with self-similar models, properties measured within a redshift-dependent density contrast are the most appropriate. In order to simplify the notation, integrated quantities with a numerical subscript, say $M_{\Delta(z)}$, refer to that quantity within a radius ($R_{\Delta(z)}$) enclosing a mean density of $\Delta(z)\rho_c(z)$. For example, $M_{200(z)}$ refers to the mass within a radius enclosing a mean overdensity of $200$ at $z=0$, and some higher mean overdensity (given by Equation \ref{c4eqn_da}) at higher redshifts. Note that this use of a redshift-dependent density contrast to define radii differs from the methods used in our previously published studies of some of these clusters. In \citet{mau04a} and \citet{mau04b}, fixed (redshift-independent) density contrasts were used, while in \citet{mau03a} the virial radii were estimated from the temperatures of the clusters. The values of integrated cluster properties quoted in this paper thus differ from those given in the previous studies.

Under the self-similar model, with the assumptions that clusters are spherically symmetric systems and that they virialised at the redshift of observation (late formation), simple scaling relations between cluster properties can be derived, based on the virial theorem \citep[\egc][]{bry98}. The total gravitating mass within a radius $R_{\Delta(z)}$ is related to the gas temperature by
\begin{eqnarray}\label{c4eqn_MT}
M_{\Delta(z)} E(z) \Delta(z)^{1/2} & \propto & kT^{3/2}
\end{eqnarray}
and the gas mass within $R_{\Delta(z)}$ is given by the relation,
\begin{eqnarray}\label{c4eqn_MgT}
M_{g\ \Delta(z)} E(z) \Delta(z)^{1/2} & \propto & kT^{3/2}\fgas.
\end{eqnarray}

If the relative distribution of gas and dark matter does not change with redshift or temperature, \ie the gas-mass fraction \fgas\ is independent of $z$ and $kT$, then Equation \ref{c4eqn_MgT} becomes similar to Equation \ref{c4eqn_MT}. This assumption is supported by measurements of \fgas\ in high-redshift clusters that are consistent with those in local clusters \citep[\egc][]{all02a,mau04a}. However, it has been found that \fgas\ is lower in clusters with temperatures below $\sim3-4\keV$ \citep[][hereafter S03]{san03}, which may invalidate the assumption of invariant \fgas\ for those systems.

Under the additional assumption that clusters emission is dominated by bremsstrahlung in the X-ray band (a reasonable assumption at the temperatures considered here), the X-ray luminosity within $R_{\Delta(z)}$ is given by 
\begin{eqnarray}\label{c4eqn_LT}
L_{\Delta(z)} E(z)^{-1} \Delta(z)^{-1/2} & \propto & kT^{2}\fgas^2.
\end{eqnarray}

The scaling between total mass and X-ray luminosity (\ML\ relation) is then given by combining Equations \ref{c4eqn_MT} and \ref{c4eqn_LT}, yielding
\begin{eqnarray}\label{c4eqn_ML}
L_{\Delta(z)} E(z)^{-7/3} \Delta(z)^{-7/6} & \propto & M_{\Delta(z)}^{4/3}\fgas^2.
\end{eqnarray}

\section{Summary of Cluster Properties and Line-Fitting Methods}
The methods used to derive the properties of the clusters in the sample can be summarised as follows. Overdensity radii $R_{\Delta(z)}$ were measured as described in \textsection \ref{c4sect_anal}. The luminosities measured within \rd\ were extrapolated to $R_{\Delta(z)}$ using the best-fitting surface-brightness profiles. The fractional detection radius $\rd/R_{200(z)}$ varied from 0.3 to 0.9, while $\rd/R_{2500(z)}$ was greater than unity for all clusters (see Table \ref{c4tab_prop}). The luminosities were thus scaled from \rd\ to $R_{200(z)}$ by factors in the range $1.0-1.5$ (median$=1.1$). 

Similarly, the gas masses were derived at different $R_{\Delta(z)}$ by extrapolation of the gas-mass profiles, obtained from the surface-brightness profiles. The extrapolation factors of \Mgas\ from \rd\ to $R_{200(z)}$ were in the range $1.1-5.3$ with a median of $1.9$. The total-mass profiles were used to derive the total masses at different overdensity radii, being scaled up by a factor in the range $1.1-3.8$ (median$=1.7$) from \rd\ to $R_{200(z)}$. 

For ClJ0046.3$+$8530 and ClJ1226.9$+$3332, temperature profiles and hardness-ratio maps support the isothermal assumption to at least $0.6R_{200(z)}$. The effect of possible departures from isothermality is investigated below. 

To enable comparisons with other work, the integrated properties of the WARPS sample are summarised in Table \ref{c4tab_summary}, derived within different radii.

\begin{table*}
\centering
\scalebox{0.9}{
\begin{tabular}{ccccccccccc} \hline 
Cluster & z & $kT$ & $R_{200(z)}$ & $L_{200(z)}$ & $M_{g\ 200(z)}$ & $M_{200(z)}$ & $R_{2500(z)}$ & $L_{2500(z)}$ & $M_{g\ 2500(z)}$ & $M_{2500(z)}$ \\
 & & $\keV$ & $\Mpc$ & $10^{44}\ergps$ & $10^{13}\Msol$ & $10^{14}\Msol$ & $\Mpc$ & $10^{44}\ergps$ & $10^{12}\Msol$ & $10^{13}\Msol$ \\ \hline

ClJ0046.3$+$8530 & 0.62 & $4.4^{+0.5}_{-0.4}$ & $0.96^{+0.06}_{-0.07}$ & $4.11^{+0.23}_{-0.19}$ & $3.69^{+0.34}_{-0.26}$ & $2.81^{+0.59}_{-0.56}$ & $0.24^{+0.02}_{-0.03}$ & $2.01^{+0.35}_{-0.29}$ & $3.74^{+0.86}_{-0.83}$ & $5.29^{+1.46}_{-1.29}$ \\

ClJ1342.9$+$2828$^\dag$ & 0.71 & $3.7^{+0.5}_{-0.4}$ & $0.88^{+0.06}_{-0.07}$ & $3.52^{+0.14}_{-0.18}$ & $3.08^{+0.32}_{-0.22}$ & $2.46^{+0.59}_{-0.52}$ & $0.18^{+0.02}_{-0.03}$ & $1.50^{+0.36}_{-0.34}$ & $2.27^{+0.83}_{-0.76}$ & $2.86^{+1.28}_{-1.00}$ \\

ClJ1113.1$-$2615 & 0.72 & $4.7^{+0.9}_{-0.7}$ & $0.97^{+0.09}_{-0.08}$ & $3.85^{+0.32}_{-0.33}$ & $2.92^{+0.35}_{-0.33}$ & $3.39^{+1.02}_{-0.78}$ & $0.26^{+0.02}_{-0.03}$ & $2.69^{+0.26}_{-0.24}$ & $4.75^{+0.94}_{-0.76}$ & $7.75^{+2.59}_{-2.15}$ \\

ClJ1103.6$+$3555$^\dag$ & 0.78 & $6.0^{+0.9}_{-0.7}$ & $0.98^{+0.07}_{-0.07}$ & $4.98^{+0.22}_{-0.21}$ & $4.06^{+0.45}_{-0.34}$ & $3.82^{+0.88}_{-0.73}$ & $0.24^{+0.02}_{-0.02}$ & $2.30^{+0.31}_{-0.30}$ & $3.92^{+0.89}_{-0.76}$ & $7.22^{+2.09}_{-1.87}$ \\

ClJ0152-7$-$1357N$^\dag$ & 0.83 & $5.6^{+1.0}_{-0.8}$ & $1.00^{+0.10}_{-0.12}$ & $11.50^{+1.52}_{-1.39}$ & $6.48^{+1.37}_{-0.75}$ & $4.32^{+1.54}_{-1.34}$ & $0.15^{+0.05}_{-0.08}$ & $2.62^{+1.58}_{-1.85}$ & $1.77^{+2.11}_{-1.49}$ & $1.90^{+2.48}_{-1.70}$ \\

ClJ0152-7$-$1357S & 0.83 & $4.8^{+1.1}_{-1.0}$ & $0.90^{+0.10}_{-0.12}$ & $6.72^{+1.07}_{-0.89}$ & $3.84^{+0.80}_{-0.62}$ & $3.14^{+1.22}_{-1.04}$ & $0.22^{+0.03}_{-0.04}$ & $3.96^{+0.57}_{-0.65}$ & $4.86^{+1.38}_{-1.49}$ & $6.16^{+3.06}_{-2.53}$ \\

ClJ1559.1$+$6353$^\dag$ & 0.85 & $4.1^{+1.4}_{-1.0}$ & $0.78^{+0.14}_{-0.12}$ & $2.59^{+0.33}_{-0.27}$ & $1.92^{+0.51}_{-0.36}$ & $2.12^{+1.40}_{-0.84}$ & $0.21^{+0.04}_{-0.04}$ & $1.68^{+0.39}_{-0.46}$ & $2.86^{+1.13}_{-0.95}$ & $5.28^{+3.72}_{-2.27}$ \\

ClJ1008.7$+$5342 & 0.87 & $3.6^{+0.8}_{-0.6}$ & $0.76^{+0.08}_{-0.09}$ & $3.84^{+0.28}_{-0.33}$ & $2.74^{+0.47}_{-0.32}$ & $2.00^{+0.81}_{-0.62}$ & $0.14^{+0.04}_{-0.05}$ & $1.16^{+0.70}_{-0.56}$ & $1.23^{+1.15}_{-0.77}$ & $1.57^{+1.56}_{-1.06}$ \\

ClJ1226.9$+$3332 & 0.89 & $10.6^{+1.1}_{-1.1}$ & $1.29^{+0.06}_{-0.08}$ & $43.70^{+0.96}_{-0.96}$ & $11.90^{+0.89}_{-0.82}$ & $10.20^{+1.71}_{-1.68}$ & $0.35^{+0.02}_{-0.03}$ & $32.70^{+1.39}_{-1.46}$ & $21.50^{+1.93}_{-2.22}$ & $25.00^{+4.61}_{-4.32}$ \\

ClJ1429.0$+$4241$^\dag$ & 0.92 & $6.2^{+1.5}_{-1.0}$ & $0.97^{+0.11}_{-0.11}$ & $9.59^{+0.92}_{-0.82}$ & $4.29^{+0.76}_{-0.60}$ & $4.49^{+1.73}_{-1.29}$ & $0.26^{+0.03}_{-0.04}$ & $6.92^{+0.70}_{-0.76}$ & $7.28^{+1.53}_{-1.57}$ & $10.50^{+4.77}_{-3.26}$ \\

ClJ1415.1$+$3612 & 1.03 & $5.7^{+1.2}_{-0.7}$ & $0.88^{+0.08}_{-0.08}$ & $10.40^{+0.62}_{-0.58}$ & $3.85^{+0.54}_{-0.43}$ & $3.83^{+1.20}_{-0.94}$ & $0.23^{+0.02}_{-0.03}$ & $7.58^{+0.70}_{-0.81}$ & $6.73^{+1.22}_{-1.19}$ & $8.82^{+3.12}_{-2.45}$ \\

\hline

\end{tabular}
}
\caption{\label{c4tab_summary}Summary of the integrated properties of the WARPS high-redshift sample derived within different radii. Luminosities are bolometric X-ray luminosities. $^\dag$These systems show evidence of being unrelaxed}
\end{table*}

The observed properties of the WARPS sample, and others, are compared with
the predicted scaling relations in the following sections. The best-fitting
scaling relations to the observations were found by performing an
orthogonal, weighted ``BCES'' regression \citep[as described by][]{akr96},
on the data in log space. This method takes into account measurement errors
on both variables, correlations in those errors, and intrinsic scatter in
the data. The use of orthogonal regression avoids the biases inherent in
bisector-regression fits to data with intrinsic scatter. The errors on the
slopes and normalisations of the relations were derived from jackknife
analyses of the datasets. The $\chisq$ goodness of fit values of the
various relations were computed including the errors in both the x and y
directions. In a similar way, the intrinsic scatter ($\sigma_s$) of the data
about the relations was defined in log space as
\begin{eqnarray}\label{eqn.scat}
\sigma_s & = & \frac{1}{N}\sum_{i=1}^N{\left[\frac{(y_i-mx_i-c)^2}{\sigma^2_{yi}+m^2\sigma^2_{xi}}\right]^{1/2}},
\end{eqnarray}
where $m$ and $c$ are the gradient and intercept of the relation, and
$(x_i,y_i)$ are the coordinates of each of the $N$ data points, with uncertainties
$(\sigma_{xi},\sigma_{yi})$. This measurement was used to compare the
scatter in the different relations discussed here.

\section{The \LT\ Relation}\label{c4sect_LT}
The effect of cool cores needs to be accounted for in any measurement of the \LT\ relation. Many relaxed clusters have dense cooling cores, with gas temperatures in the central $\sim100\kpc$ falling to $\sim1/3$ of the global temperature, and sharply peaked surface-brightness profiles \citep[\egc][]{fab94b,kaa01}. These effects are not included in the self-similar model, so must be taken into account in the analysis. This is routinely done by excluding the central region of such clusters from the analyses. None of the clusters in our sample showed significant evidence for cooling cores; where temperature profiles and hardness-ratio maps could be created (ClJ0046.3$+$8530 and ClJ1226.9$+$3332) there was no indication of cooler gas in the central regions, and none of the surface-brightness profiles were centrally peaked. For this reason, no cooling-core correction was applied to the luminosities or temperatures of the high-redshift clusters.

\begin{figure*}
\begin{center}
\scalebox{0.55}{\includegraphics*[angle=270]{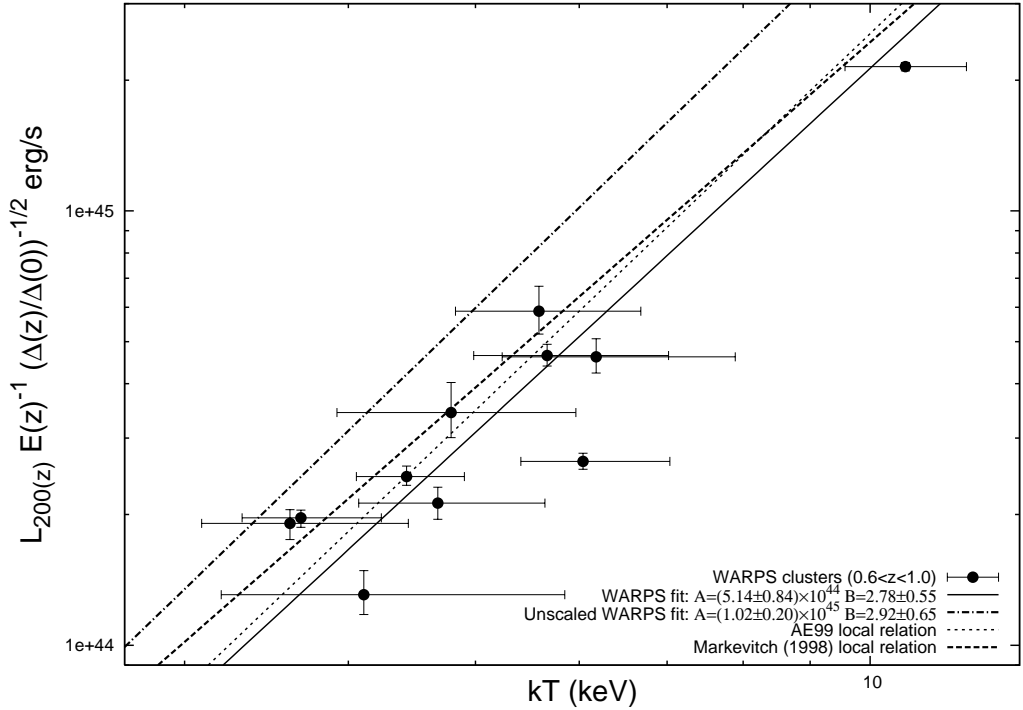}} \\
\caption[\LT\ relation for the WARPS sample.]{\label{c4fig_myLT}\LT\
relation for the high-redshift WARPS sample. Luminosities were extrapolated
to a radius $R_{200(z)}$, corresponding to a redshift-dependent density
contrast $200$, and scaled by a factor of
$E(z)^{-1}[\Delta(z)/\Delta(0)]^{-1/2}$ as predicted by the self-similar
model. The solid line is the best fit to the data, and the dotted and
dashed lines are the local relations of AE99 and \citet{mar98a}
respectively. The dot-dashed line is the best fit to the unscaled
high-redshift clusters (points not plotted) and can be used to judge the
significance of the self-similar scaling.}  
\end{center}
\end{figure*}

If the \LT\ relation evolves as predicted by Equation \ref{c4eqn_LT}, then
the effect of evolution can be removed by dividing the luminosity of each
cluster by $E(z)^{-1}(\Delta(z)/\Delta(0))^{-1/2}$, which reduces to
$E(z)^{-1}(\Dv(z)/\Dv(0))^{-1/2}$. Following this scaling, the high-z
clusters should lie on the local relation. Of the two factors describing the evolution, $E(z)$ dominates; at $z=1$, $E(z)^{-1}=0.57$ and $(\Dv(z)/\Dv(0))^{-1/2}=0.80$. Fig. \ref{c4fig_myLT} shows the scaled luminosities (those with self-similar evolution factored out) measured within $R_{200(z)}$ plotted against temperature. A relation of the form 
\begin{eqnarray}
E(z)^{-1}\left(\frac{\Dv(z)}{\Dv(0)}\right)^{-1/2}L_{\Delta(z)} & = & A \left(\frac{kT}{6\keV}\right)^B
\end{eqnarray}
was fit to the data, resulting in best-fit values
$A=(5.14\pm0.84)\times10^{44}h_{70}^{-2}\ergps$, $B=2.78\pm0.55$. The
dot-dashed line in Fig. \ref{c4fig_myLT} shows the best-fitting relation
with the \emph{observed} luminosities, illustrating the effect of the
predicted evolution.

Also plotted on Fig. \ref{c4fig_myLT} is the local \LT\ relation measured
by \citet[][hereafter AE99]{arn99} for a sample of clusters with little or
no central cooling. It is unclear within what precise radius the AE99
luminosities were determined, however as they are referred to as ``total
luminosities'' we assume they are derived within $R_{200(z)}$. If the AE99
luminosities were extrapolated to some larger radius (or to infinity), then
the luminosities will be $\lta10\%$ higher than the values within
$R_{200(z)}$. The best fit to our scaled high-redshift data is consistent,
within the errors, with the parameters for the local relation of
$A=(5.86\pm0.40)\times10^{44}h_{70}^{-2}$, $B=2.88\pm0.15$. The local AE99
relation provides an acceptable fit to the scaled high-redshift data
($\chisq/\nu=10.2/9$), while the unscaled high-redshift data rule out the
local relation at $>95\%$ level($\chisq/\nu=18.9/9$).

The scaled WARPS high-z \LT\ relation is also compared with the local
relation measured by \citet{mar98a} in Fig. \ref{c4fig_myLT}. In contrast
with AE99, \citet{mar98a} corrected for central cooling by excising the
central regions of the clusters and extrapolating a surface-brightness
profile over this region. The luminosities measured by \citet{mar98a} were
extrapolated to a fixed radius of $1.4h_{70}^{-1}\Mpc$, so are expected to
be $\approx2\%$ higher than the values derived within $R_{200(z)}$ (based
on the typical scaling of luminosity from $1.4h_{70}^{-1}\Mpc$ to
$R_{200(z)}$ for our sample). The best-fitting relation found by
\citet{mar98a} has $A=(6.35\pm0.55)\times10^{44}h_{70}^{-2}$ and
$B=2.64\pm0.27$ in our notation. While the WARPS data are also reasonably
consistent with this local relation ($\chisq/\nu=14.2/9$), we prefer to
compare the results with AE99 because the treatment of central cooling is
consistent. 

\subsection{Comparison with \citet{vik02}}\label{sect_v02}
The results for the WARPS sample was compared with those obtained by
\citet[][hereafter V02]{vik02} for a sample consisting of 22 clusters at
$z>0.4$ observed with \Chandra. In the V02 sample, luminosities were
extrapolated to a fixed radius of $1.4h_{70}^{-1}\Mpc$ irrespective of
their redshift. In order to compare the V02 luminosities with self-similar
predictions, they were scaled to $R_{200(z)}$. The value of $R_{200(z)}$
was computed in an identical way to the WARPS clusters (see \textsection
\ref{c4sect_anal}) using the values of $z$, $kT$, $\rc$ and $\beta$ given
by V02. Random realisations of the overdensity profiles were computed from
the uncertainties on $z$, $kT$, $\beta$, given by V02. The luminosity of
each system was then scaled from $1.4h_{70}^{-1}\Mpc$ to $R_{200(z)}$ based
on $\beta$-profiles with the parameters given by V02 for that cluster. The
resulting scale factors for the luminosities were close to unity, ranging
from $0.89$ to $1.02$ with a mean of $0.98$. 

As no errors on $\Lx$ are given by V02, the mean fractional uncertainty
from the WARPS sample of $0.09$ was assumed for the V02 clusters. The
errors on the luminosities scaled to $R_{200(z)}$ thus include the
statistical uncertainties in modeling the overdensity profiles, and an
additional $9\%$ uncertainty. 

\begin{figure*}
\begin{center}
\scalebox{0.55}{\includegraphics*[angle=270]{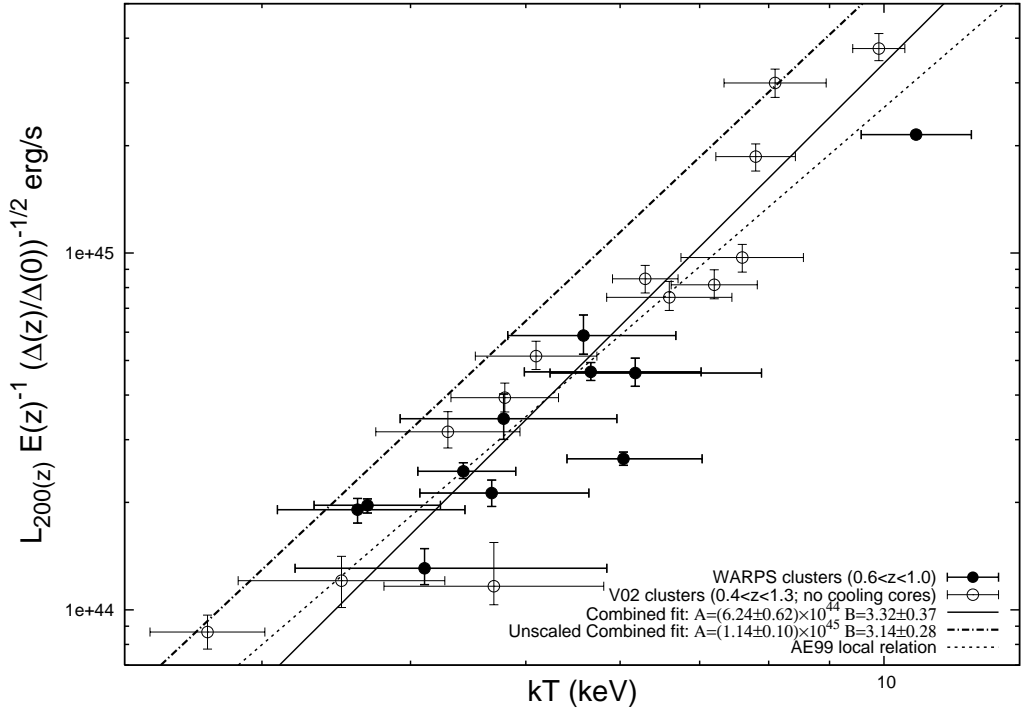}} \\
\caption[\LT\ relation for the combined WARPS and V02
samples.]{\label{c4fig_v02LT}\LT\ relation for the combined WARPS and V02
samples. The WARPS data are the same as in Fig. \ref{c4fig_myLT}. The V02
data were treated in the same way, with \Lx\ extrapolated to $R_{200(z)}$
and scaled by $E(z)^{-1}[\Delta(z)/\Delta(0)]^{-1/2}$. The solid line is
the best fit to the combined dataset.} 
\end{center}
\end{figure*}

V02 excluded the central $71h_{70}^{-1}\kpc$ of clusters which had peaked
surface-brightness profiles, and extrapolated over that region to correct
for central cooling effects. As no cooling corrections were applied to the
WARPS clusters, the corrected clusters were discarded from the V02 sample
(this is a fairly conservative move as the corrected clusters do not
scatter from the V02 \LT\ relation). We also excluded ClJ0152.7$-$1357,
which is already in the WARPS sample, leaving 13 V02 clusters. The
best-fitting \LT\ relation for those clusters was consistent with the WARPS
\LT\ relation shown in Fig. \ref{c4fig_myLT}. The two samples were then
combined to give a sample of 22 systems, with 16 at $z>0.6$. The
luminosities were scaled as before to remove the predicted self-similar
evolution, and the resulting \LT\ relation is shown in
Fig. \ref{c4fig_v02LT}, along with the local AE99 relation. 

The best-fitting parameters for the joint sample are
$A=(6.24\pm0.62)\times10^{44}h_{70}^{-2}$ and $B=3.32\pm0.37$. This is
marginally consistent with the local AE99 relation, with
$\chisq/\nu=33.3/22$. However, this statistic does not take into account
any intrinsic scatter in the data so the null hypothesis probability of
$6\%$ is an underestimate to some extent. We thus conclude that the scaled
high-redshift data cannot exclude the local AE99 relation as an acceptable
model. Comparison of the \emph{unscaled} data for the combined WARPS and
V02 samples (indicated by the dot-dashed line in Fig. \ref{c4fig_v02LT})
with the AE99 relation strongly rules out the local relation as a
description of the high-redshift data ($\chisq/\nu=130/22$).  

As the WARPS results are based mainly on \XMM\ observations, a potential
source of systematic error is that cooling cores in high-redshift systems
could go undetected due to the large PSF of \XMM. However, the best-fitting
relation for the combination of the WARPS sample and the V02 data
(excluding the systems with cooling cores detected by \Chandra) is a good
description of both samples (Fig. \ref{c4fig_v02LT}). Both samples also
show a similar amount of scatter about the best-fitting relation. This
suggests that the \XMM\ measurements were not contaminated by undetected
cooling cores, and that the cross-calibration of the instruments is not a
significant concern here.

\subsection{Comparison with \citet{ett04}}
\begin{figure*}
\begin{center}
\scalebox{0.55}{\includegraphics*[angle=270]{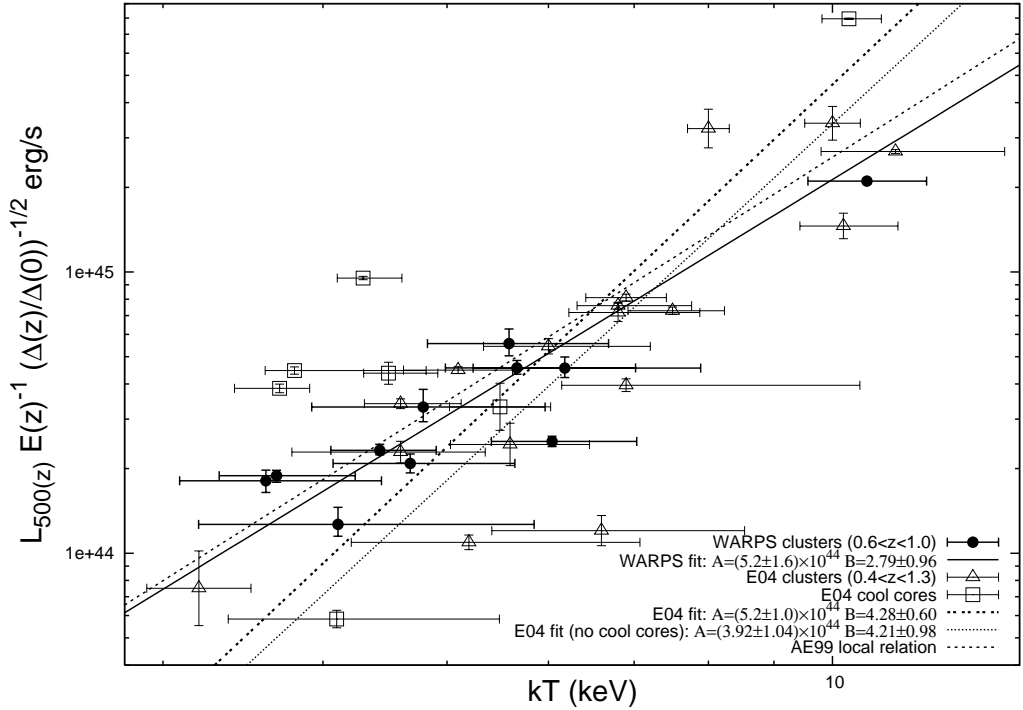}} \\
\caption[\LT\ relation for the WARPS and E04
samples.]{\label{c4fig_e04LT}\LT\ relations of the WARPS and E04
samples. Luminosities were extrapolated to $R^{\prime}_{500(z)}$.} 
\end{center}
\end{figure*}

The WARPS sample was then compared with the larger, more recently compiled sample of \citet[][hereafter E04]{ett04} which comprises 28 clusters at $z>0.4$ from the \Chandra\ archive. In E04, luminosities were extrapolated to $R^{\prime}_{500(z)}$, where the prime indicates a slightly different definition of $\Delta(z)$ used in that work. The density contrast in E04 was defined as
\begin{eqnarray}\label{c4eqn_dap}
\Delta^{\prime}(z)=500\frac{\Dv(z)}{18\pi^2},
\end{eqnarray}
so is related to our $\Delta(z)$ by
\begin{eqnarray}\label{c4eqn_dap2}
\Delta^{\prime}(z)=\frac{\Dv(0)}{18\pi^2}\Delta(z).
\end{eqnarray}
In the case of an Einstein-de Sitter universe
$\Delta^{\prime}(z)=\Delta(z)$. In our assumed low-density \LCDM cosmology
however, $\Dv(0)/18\pi^2=1.76$ and so $\Delta^{\prime}(z)<\Delta(z)$,
leading to higher measured luminosities within $R^{\prime}_{500(z)}$. This
change in definition thus introduces a redshift-independent change in all
luminosities, but the predicted evolution of the \LT\ relation is
unaffected. For consistency with E04, the luminosities of the WARPS
clusters were extrapolated to $R^{\prime}_{500(z)}$. The resulting WARPS
\LT\ relation is plotted along with the E04 relation in
Fig. \ref{c4fig_e04LT}. The AE99 relation is also plotted for comparison,
though it should be recalled that those luminosities were extrapolated to
larger radii. In addition to any evolution, the normalisation of the AE99
relation should thus be $\approx5\%$ higher than the $R^{\prime}_{500(z)}$
relations (estimated from the extrapolation of a standard $\beta$-profile
from $R^{\prime}_{500(z)}$ to $R_{200(z)}$). 

The best fit to the WARPS $R^{\prime}_{500(z)}$ \LT\ relation is consistent
with the $R_{200(z)}$ relation of Fig. \ref{c4fig_myLT}, and the local AE99
relation (see Table \ref{c4tab_LT}). The best fit to the E04 relation, with
$A=(5.2\pm1.1)\times10^{44}h_{70}^{-2}$ and $B=4.28\pm0.60$, is possibly
steeper than the other \LT\ relations discussed here (at the $\sim2\sigma$
level). We note that the slope we measure for the E04 sample is steeper
than that quoted in E04 ($B=3.72\pm0.47$). This is because we have scaled
the luminosities by $E(z)^{-1}(\Dv(z)/\Dv(0))^{-1/2}$ whereas E04 scaled by
$E(z)^{-1}$ alone, and because we use an orthogonal BCES regression, while
E04 used a bisector BCES regression. If we eliminate these two differences,
we recover the same fit as E04. 

The scatter of the E04 data about the best-fit \LT\ relation is
$\sigma_s=1.78$ which is larger than that of the combined WARPS and V02
sample ($\sigma_s=0.85$; \cf Figs. \ref{c4fig_v02LT} and
\ref{c4fig_e04LT}). One probably cause of this is that E04 made no
correction for cool cores, while 7 of the clusters were classed as having
cool cores by V02. These cool core clusters are indicated in
Fig. \ref{c4fig_e04LT}, and their exclusion reduces both the scatter
($\sigma_s=1.25$) and also the normalisation of the best fitting
relation. In light of these differences, and the good agreement between the
WARPS and V02 \LT\ relations, we prefer to use the combined WARPS and V02
\LT\ relation, with luminosities measured within $R_{200(z)}$ in our
further discussions. The different \LT\ relations discussed in this section
are summarised in Table \ref{c4tab_LT}. 

\begin{table*}
\centering
\scalebox{1.0}{
\begin{tabular}{llll} \hline 
\LT\ relation & $A$ ($10^{44}h_{70}^{-2}\ergps$) & $B$ & notes \\
\hline \hline

\multicolumn{4}{l}{\bf Luminosities extrapolated to $R_{200(z)}$}\\

AE99 & $5.86\pm0.40$ & $2.88\pm0.15$ & Local clusters with no strong central cooling (Fig. \ref{c4fig_myLT}).\\

\citet{mar98a} & $6.35\pm0.55$ & $2.64\pm0.27$ & Local relation, corrected for cooling cores (Fig. \ref{c4fig_myLT}). \\

WARPS  & $5.14\pm0.84$ & $2.78\pm0.55$ & $0.6<z<1.0$ clusters with no cooling cores (Fig. \ref{c4fig_myLT}).\\

V02 \& WARPS & $6.24\pm0.62$ & $3.32\pm0.37$ & Combined samples ($0.4<z<1.3$), no cooling cores (Fig. \ref{c4fig_v02LT}).\\ 

\hline

\multicolumn{4}{l}{\bf Luminosities extrapolated to $R^{\prime}_{500(z)}$}\\

E04 & $5.21\pm1.05$   & $4.28\pm0.60$ & Full E04 $0.4<z<1.3$ sample (Fig. \ref{c4fig_e04LT}). \\

E04 (no cool cores) & $3.92\pm1.04$   & $4.21\pm0.98$ & E04 $0.4<z<1.3$
clusters with no cooling cores (Fig. \ref{c4fig_e04LT}).\\

WARPS ($R^{\prime}_{500(z)}$) & $4.97\pm0.80$ & $2.80\pm0.55$ & WARPS $0.6<z<1.0$ clusters with no cooling cores (Fig. \ref{c4fig_e04LT}).\\

\hline
\end{tabular}
}
\caption[Summary of the \LT\ relations discussed in \textsection
\ref{c4sect_LT}]{\label{c4tab_LT}Summary of the \LT\ relations discussed in
\textsection \ref{c4sect_LT}. Luminosities of the high-redshift clusters
were scaled by $E(z)^{-1}(\Dv(z)/\Dv(0))^{-1/2}$ to remove the predicted
self-similar evolution.} 
\end{table*}

\section{The \MT\ Relation}\label{c4sect_MT}
In the self-similar evolution scenario, assuming the late-formation
approximation, the \MT\ relation is given by Equation \ref{c4eqn_MT}. The
evolution of the \MT\ relation was investigated by comparing the WARPS
sample to the local sample of S03, which comprises $66$ clusters with
reliable masses derived from temperature and surface-brightness
profiles. We chose S03 for this comparison because the overdensity radii
and mass measurements were made in the same way as in this
work. Importantly, S03 also derived total and gas masses for all of the
clusters in their sample under the assumption of isothermality. They found
that assuming isothermality for their entire sample, including genuinely
isothermal, and non-isothermal clusters lead to an average $\sim30\%$
overestimate of the true mass (the mass estimated with full temperature
profiles) within $R_{200}$.

This is a key point in this analysis. In all but a few cases, departures
from isothermality in the high-z clusters cannot be detected. In the cases
where there is evidence in support of isothermality, the constraints are
not strong. While some of the high-z systems may be genuinely isothermal,
others will surely not be. Assuming isothermality for the whole sample (while
correct for some clusters) is likely to lead to an {\it average} systematic
mass overestimate similar to that found by S03. In order to separate this
systematic effect from any real evolution in the \MT\ relation, it is
essential that isothermal masses be used for the low-redshift clusters.

For consistency with the WARPS sample, we include only the 40 clusters from
the S03 sample with $kT>3\keV$. The slopes and normalisations of the
relations we fit to the S03 data vary from those reported in S03 because of
this temperature cut-off, and because we use a slightly different fitting
algorithm than S03.  

A reliable masses measurement requires that the clusters be in hydrostatic
equilibrium. The five possibly unrelaxed high-redshift clusters
(ClJ1342.9$+$2828, ClJ1103.6$+$3555, ClJ0152.7$-$1357N, ClJ1559.1$+$6353
and ClJ1429.0$+$4241) were flagged in this analysis, and all of the \MT\
relations were fit with and without these systems. Excluding the unrelaxed
systems had no significant effect on any of the best fit relations, and so
they were retained to avoid biasing the sample.

Masses were derived within $R_{2500(z)}$, which falls within the detection
radius for all of the WARPS clusters, and $R_{200(z)}$, which corresponds
to the estimated virial radius used in many other studies. Similarly to the
method used with the \LT\ relation, the predicted self-similar evolution
was factored out of the high-redshift cluster masses by fitting a relation
of the form 
\begin{eqnarray}
E(z)\left(\frac{\Dv(z)}{\Dv(0)}\right)^{1/2}M_{\Delta(z)} & = & A
\left(\frac{kT}{6\keV}\right)^B 
\end{eqnarray}
to the data.

\begin{figure*}
\begin{center}
\scalebox{0.55}{\includegraphics*[angle=270]{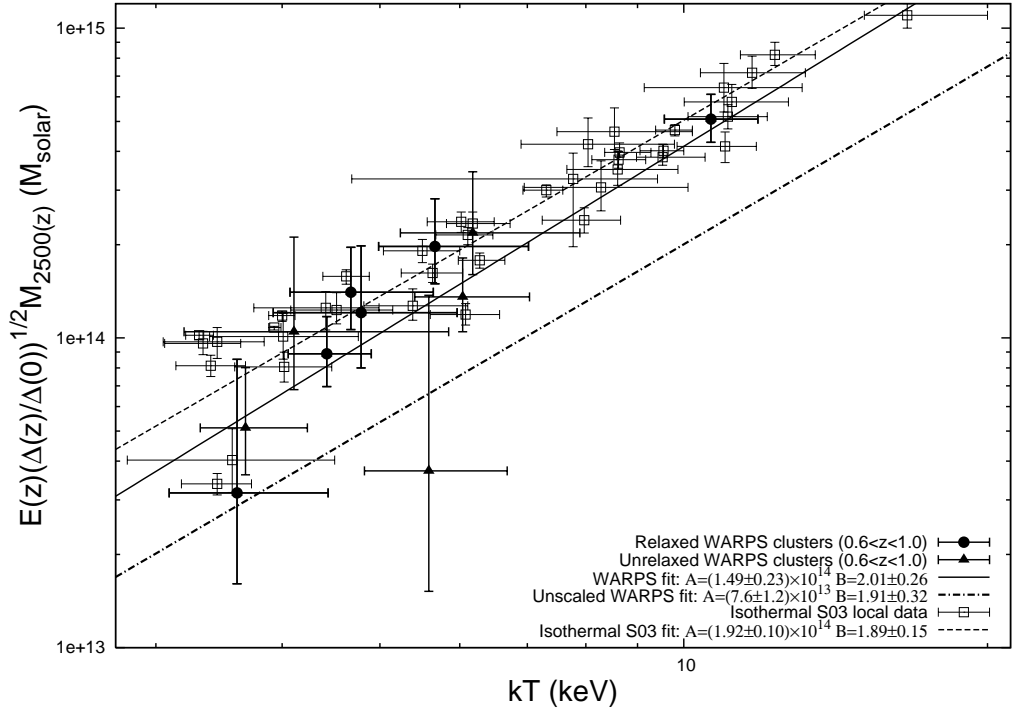}} \\
\caption[\MnT{2500(z)}\ relation for the low-z S03 clusters and the relaxed
high-z WARPS systems.]{\label{c4fig_MT2500}\MnT{2500(z)}\ relation for the
low-z S03 clusters and the relaxed high-z WARPS systems. Masses were
measured within $R_{2500(z)}$ assuming isothermality and scaled by the
evolution predicted by the self-similar model. The dot-dashed
line is the best fit to the unscaled high-redshift clusters (points not
plotted) and can be used to judge the significance of the self-similar scaling.}
\end{center}
\end{figure*}

Fig. \ref{c4fig_MT2500} shows the \MnT{2500(z)}\ relation for the S03 and
WARPS samples. The masses for both samples were derived assuming
isothermality. The temperatures of the S03 clusters are emission-weighted
temperatures measured within $0.3R_{200(z)}$, which corresponds closely to
$R_{2500(z)}$, and are extrapolated over any central cool gas (see
S03). The WARPS temperatures are also emission weighted, and are measured
within the clusters' detection radii. These measurements are consistent,
under the assumption of isothermality, with the S03 temperature
measurements. Our best-fitting relation for the S03 data is given by
$A=(1.92\pm0.10)\times10^{14}h_{70}$ and $B=1.89\pm0.15$, while the
best-fit to the WARPS data is parameterised by
$A=(1.49\pm0.23)\times10^{14}h_{70}$ and $B=2.01\pm0.26$.  

The \MnT{2500(z)}\ relation was also measured for the WARPS systems when
their masses were not scaled by the predicted evolution, and is plotted as
a dot-dashed line in Fig. \ref{c4fig_MT2500}. These unscaled masses provide
strong evidence for evolution of the \MnT{2500(z)}\ relation; the $\chisq$
of the local relation to the unscaled high-redshift data is
$\chisq/\nu=37.3/9$. The \emph{scaled} high-redshift data agree well with
the local isothermal relation, with $\chisq/\nu=4.4/9$. 

\begin{figure*}
\begin{center}
\scalebox{0.55}{\includegraphics*[angle=270]{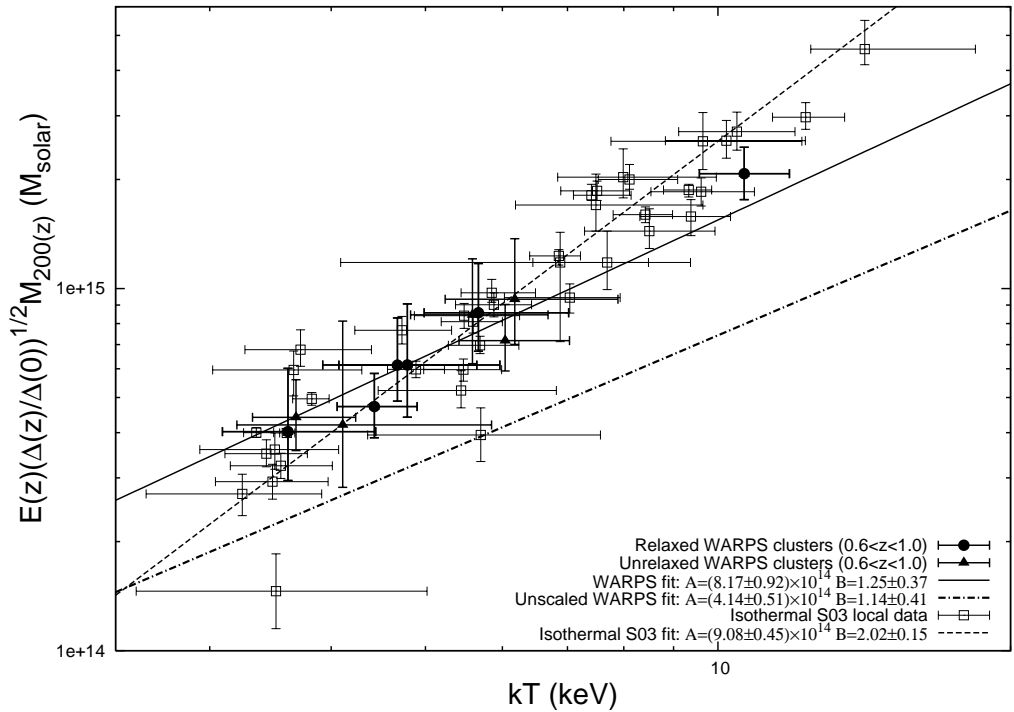}} \\
\caption[\MnT{200(z)}\ relation for the low-z S03 clusters and the relaxed
high-z WARPS systems.]{\label{c4fig_MT200}\MnT{200(z)}\ relation for the
low-z S03 clusters and the relaxed high-z WARPS systems. Masses were
measured within $R_{200(z)}$ assuming isothermality and scaled by the
evolution predicted by the self-similar model. The dot-dashed line is the
best fit to the unscaled high-redshift clusters (points not plotted) and
can be used to judge the significance of the self-similar scaling.} 
\end{center}
\end{figure*}

The \MnT{200(z)}\ relation is shown in Fig. \ref{c4fig_MT200} for the S03
and WARPS samples. The temperatures of the S03 clusters are as above, but
measured within $R_{200(z)}$, while the WARPS temperatures are again
measured within \rd. The isothermal S03 masses were used and the best-fit
parameters are summarised in Table \ref{c4tab_MT}. Again, the scaled
high-redshift data are consistent with the local isothermal relation
($\chisq/\nu=2.9/9$). The best-fitting relation for the WARPS data using
masses not scaled by the predicted evolution is also plotted in
Fig. \ref{c4fig_MT200} as a dot-dashed line. The unscaled high-redshift
data are inconsistent with the local isothermal \MnT{200(z)}\ relation
($\chisq/\nu=38.8/9$), providing evidence for evolution.

To enable comparisons with other work, masses of the WARPS clusters were
also derived within radii enclosing fixed, redshift-independent density
contrasts of $\Delta(z)=\Delta(0)=2500$ and
$\Delta(z)=\Delta(0)=200$. These data were then fit with a relation of the
form 
\begin{eqnarray}
M_{\Delta(z)} & = & A \left(\frac{kT}{6\keV}\right)^B,
\end{eqnarray}
so no scaling for the predicted self-similar evolution was made. The
best-fitting slopes and normalisations are given in Table \ref{c4tab_MT}. 

\begin{table*}
\centering
\scalebox{1.0}{
\begin{tabular}{llll} \hline 
\MT\ relation & $A$ ($10^{14}h_{70}^{-1}\Msol$) & $B$ & notes \\
\hline \hline

\multicolumn{4}{l}{\bf Masses within $R_{2500(z)}$}\\

WARPS \MnT{2500(z)}\ & $1.49\pm0.23$ & $2.01\pm0.26$ & $0.6<z<1.0$ clusters,
isothermal masses (Fig. \ref{c4fig_MT2500}).\\

S03 \MnT{2500(z)}\ & $1.92\pm0.10$ & $1.89\pm0.15$ & Local clusters,
isothermal masses (Fig. \ref{c4fig_MT2500}).\\

\hline

\multicolumn{4}{l}{\bf Masses within $R_{200(z)}$}\\

WARPS \MnT{200(z)}\ & $8.17\pm0.92$ & $1.25\pm0.37$ & $0.6<z<1.0$ clusters,
isothermal masses (Fig. \ref{c4fig_MT200}).\\

S03 \MnT{200(z)}\ & $9.08\pm0.45$ & $2.02\pm0.15$ & Local clusters,
isothermal masses (Fig. \ref{c4fig_MT200}).\\

\hline

\multicolumn{4}{l}{\bf Unscaled masses within redshift-independent density contrast}\\

WARPS \MnT{2500}\ & $1.10\pm0.19$ & $1.40\pm0.53 $ & $0.6<z<1.0$ clusters,
isothermal masses.\\

WARPS \MnT{200}\ & $5.14\pm0.61$ & $1.16\pm0.39$ & $0.6<z<1.0$ clusters,
isothermal masses.\\

\hline

\end{tabular}
}
\caption[Summary of the \MT\ relations discussed in \textsection
\ref{c4sect_MT}]{\label{c4tab_MT}Summary of the \MT\ relations discussed in
\textsection \ref{c4sect_MT}. Masses of the high-redshift clusters were
scaled by $E(z)(\Dv(z)/\Dv(0))^{1/2}$ to remove the predicted self-similar
evolution, with the exception of the bottom section of the table.} 
\end{table*}

\section{The \MgT\ Relation}\label{c4sect_MgT}
The comparison of the WARPS and S03 samples also enabled the investigation
of the evolution of the \MgT\ relation. The gas mass is less dependent on
uncertainties in the cluster temperature structure than the total
mass. This is because the gas density profile is obtained easily from the
X-ray surface-brightness profile, and the gas luminosity depends strongly
on density (as $\rho^2$) and weakly on temperature (as $T^{1/2}$). This
makes the \MgT\ relation potentially a more reliable method of exploring
cluster evolution. An incorrect assumption of isothermality would still
introduce systematic effects because the definition of overdensity radii
depends on the total mass profile. The self-similar \MgT\ evolution
prediction (Equation \ref{c4eqn_MgT}) is also subject to additional
assumptions about the relative distributions of the gas and dark matter
(\textsection \ref{c4sect_theory}).

\begin{figure*}
\begin{center}
\scalebox{0.55}{\includegraphics*[angle=270]{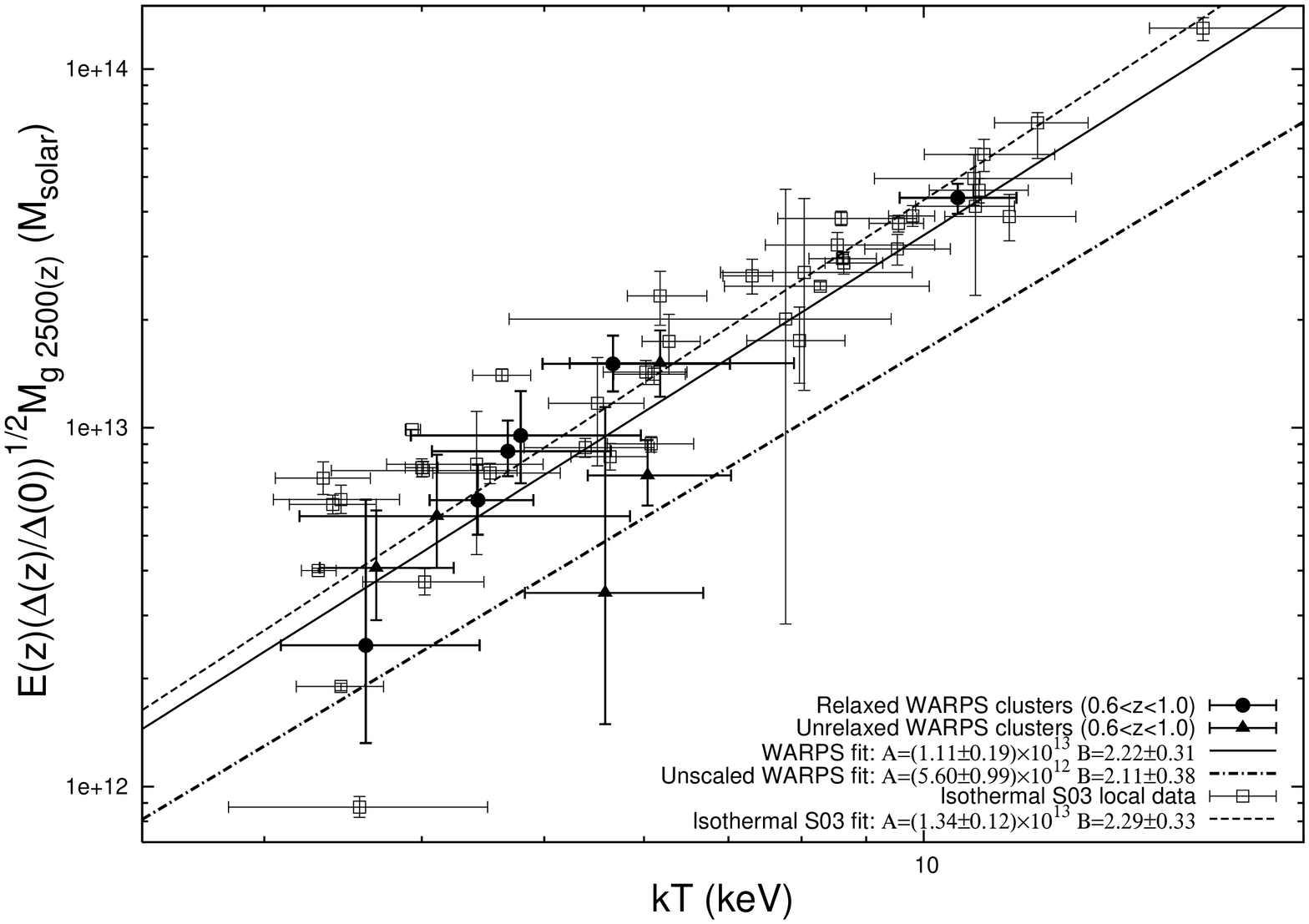}}\\
\caption[The \MnT{g\ 2500(z)}\ relations of the low-z S03 clusters and
relaxed, high-z WARPS systems.]{\label{c4fig_MgT2500}The \MnT{g\ 2500(z)}\
relations of the low-z S03 clusters and relaxed, high-z WARPS
systems. Masses were measured within $R_{2500(z)}$ assuming isothermality,
and scaled by the evolution predicted by the self-similar model. The
dot-dashed line is the best fit to the unscaled high-redshift clusters
(points not plotted) and can be used to judge the significance of the
self-similar scaling.} 
\end{center}
\end{figure*}

\begin{figure*}
\begin{center}
\scalebox{0.55}{\includegraphics*[angle=270]{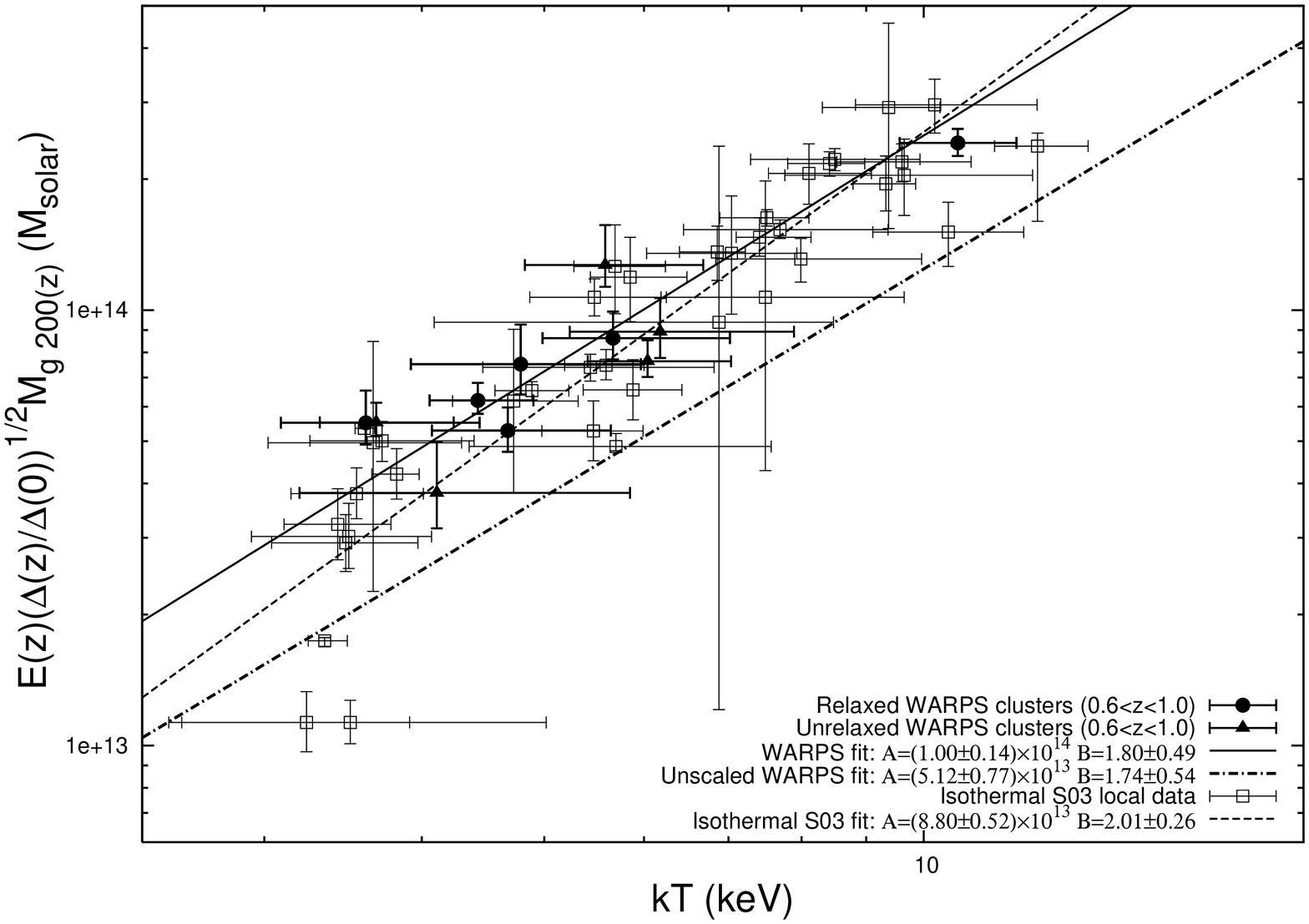}} \\
\caption[The \MnT{g\ 200(z)}\ relations of the low-z S03 clusters and
relaxed, high-z WARPS systems.]{\label{c4fig_MgT200}The \MnT{g\ 200(z)}\
relations of the low-z S03 clusters and relaxed, high-z WARPS
systems. Masses were measured within $R_{200(z)}$ assuming isothermality,
and scaled by the evolution predicted by the self-similar model.  The
dot-dashed line is the best fit to the unscaled high-redshift clusters
(points not plotted) and can be used to judge the significance of the
self-similar scaling.} 
\end{center}
\end{figure*}

The \MgT\ relations of the S03 and WARPS samples were derived within
$R_{2500(z)}$ and $R_{200(z)}$, and are plotted in
Figs. \ref{c4fig_MgT2500} and \ref{c4fig_MgT200} respectively. The
predicted self-similar evolution was again factored out of the
high-redshift cluster gas masses, with a relation of the form 
\begin{eqnarray}
E(z)\left(\frac{\Dv(z)}{\Dv(0)}\right)^{1/2}M_{g\ \Delta(z)} & = & A
\left(\frac{kT}{6\keV}\right)^B 
\end{eqnarray}
fit to the data. As with the \MT\ relation, the unrelaxed clusters in
our sample were flagged in this analysis, and the results were found to be
independent of their inclusion or exclusion. \MgT\ relations were fit to
the S03 data, using their gas masses derived under the assumption of
isothermality. Table \ref{c4tab_MgT} summarises the parameters of the
various best-fitting relations. 

Measured within $R_{2500(z)}$, the scaled high-redshift and low-redshift
\MgT\ relations are consistent, with $\chisq/\nu=4.8/9$. The scaled
high-redshift \MgT\ relation within $R_{200(z)}$ is also consistent with
the local S03 relation ($\chisq/\nu=8.0/9$). 

The \MgT\ relations for the WARPS clusters when the masses were not scaled
by the predicted self-similar evolution are plotted as dot-dashed lines in
Figs. \ref{c4fig_MgT2500} and \ref{c4fig_MgT200}. Within both radii, the
unscaled high-redshift data are strongly inconsistent with the local
isothermal relations, demonstrating the evolution of the relations.

\begin{table*}
\centering
\scalebox{1.0}{
\begin{tabular}{llll} \hline 
\MgT\ relation & $A$ ($10^{13}h_{70}^{-1}\Msol$) & $B$ & notes \\
\hline \hline

\multicolumn{4}{l}{\bf Gas masses within $R_{2500(z)}$}\\

WARPS \MnT{g\ 2500(z)}\ & $1.11\pm0.19$ & $2.22\pm0.31$ & $0.6<z<1.0$
clusters, isothermal masses (Fig. \ref{c4fig_MgT2500}).\\

S03 \MnT{g\ 2500(z)}\ & $1.34\pm0.12$ & $2.29\pm0.33$ & Local clusters, isothermal masses (Fig. \ref{c4fig_MgT2500}).\\

\hline

\multicolumn{4}{l}{\bf Gas masses within $R_{200(z)}$}\\

WARPS \MnT{g\ 200(z)}\ & $10.0\pm1.4$ & $1.80\pm0.49$ & $0.6<z<1.0$
clusters, isothermal masses (Fig. \ref{c4fig_MgT200}).\\

S03 \MnT{g\ 200(z)}\ & $8.80\pm0.52$ & $2.01\pm0.26$ & Local clusters, isothermal masses (Fig. \ref{c4fig_MgT200}).\\

\hline
\end{tabular}
}
\caption[Summary of the \MgT\ relations discussed in \textsection
\ref{c4sect_MgT}]{\label{c4tab_MgT}Summary of the \MgT\ relations discussed
in \textsection \ref{c4sect_MgT}. Masses of the high-redshift clusters were
scaled by $E(z)(\Dv(z)/\Dv(0))^{1/2}$ to remove the predicted self-similar
evolution.}
\end{table*}

V02 also investigated the evolution of the \MgT\ relation, measuring gas
masses within an overdensity radius defined in terms of the average baryon
density of the universe. Although their measurements are not directly
comparable to ours for this reason, they find evidence for weak evolution
(with respect to the local relation) in the \MgT\ relation that is
qualitatively consistent with our results.

\section{The \ML\ relation}\label{c4sect_ML}
The scaling between total mass and X-ray luminosity in the high-redshift
clusters was then investigated. The WARPS sample was compared with the
low-redshift HIFLUGCS sample of \citet{rei02} in order to measure any
evolution in the \ML\ relation. The masses and luminosities published in
\citet{rei02} were scaled to $H_0=70\kmpspMpc$, and clusters with
$kT<3\keV$ were removed for consistency with the WARPS clusters, leaving
52. The HIFLUGCS sample includes systems regardless of their morphology,
although any strong substructure was excluded for the mass and luminosity
determinations, and all of the high-redshift clusters (relaxed and
unrelaxed) are included in this comparison. In addition, the masses of the
local systems were determined assuming isothermality, which enables a fair
comparison with the high-redshift masses. Properties extrapolated to \rt
were used, and a relation of the form  
\begin{eqnarray}
E(z)^{-7/3}\left(\frac{\Dv(z)}{\Dv(0)}\right)^{-7/6}L_{200(z)} & = & A \left(\frac{M_{200(z)}}{5\times10^{14}h_{70}^{-1}\Msol}\right)^B
\end{eqnarray}
was fit to high- and low-redshift data separately.

\begin{figure*}
\begin{center}
\scalebox{0.55}{\includegraphics*[angle=270]{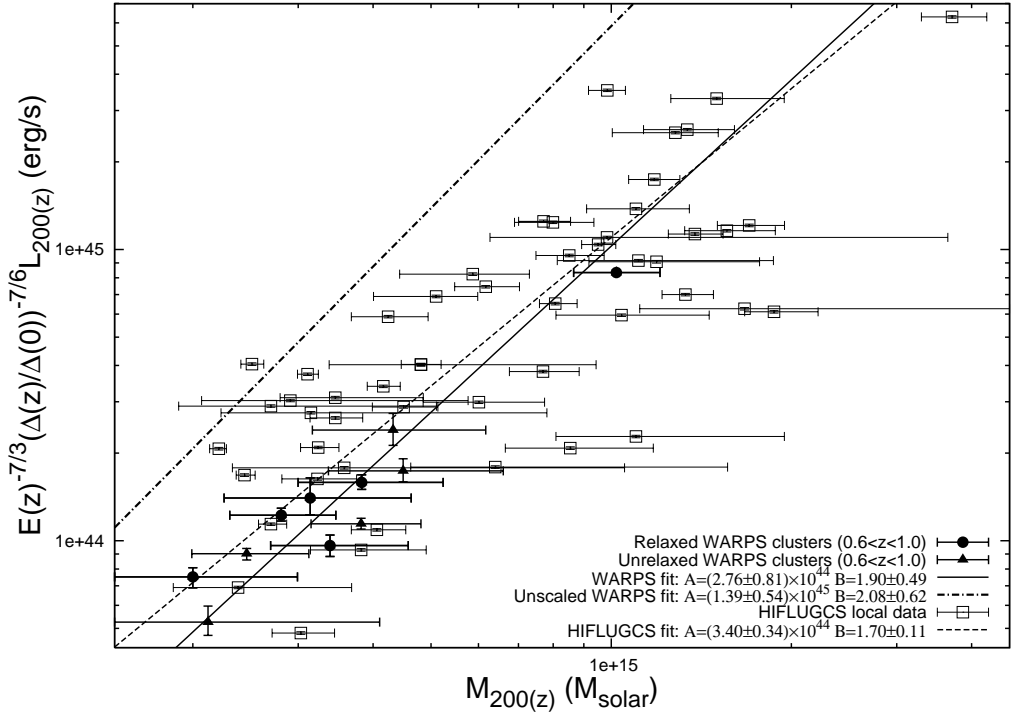}} \\
\caption[\ML\ relation for the low-z HIFLUGCS clusters and the high-z WARPS
systems.]{\label{c4fig_ML}\ML\ relation for the low-z HIFLUGCS clusters and
the high-z WARPS systems. Masses were measured within $R_{200(z)}$ assuming
isothermality and luminosities within the same radius were scaled by the
evolution predicted by the self-similar model.  The dot-dashed line is the
best fit to the unscaled high-redshift clusters (points not plotted) and
can be used to judge the significance of the self-similar scaling.}
\end{center}
\end{figure*}

Fig. \ref{c4fig_ML} shows the best-fitting \ML\ relations for the two
samples, and the parameters of the different relations are summarised in
Table \ref{c4tab_ML}. The dot-dashed line shows the best-fitting \ML\
relation when no scaling for the predicted evolution is applied. The
disagreement between that unscaled high-z relation and the local data is
striking, even given the large scatter in the local data. However, when the
high-redshift data are scaled by the predicted evolution, the best-fitting
relation (solid line) agrees very well with the local relation (dashed
line), with $\chisq/\nu=9.3/9$. The slopes of the low- and high-z relations
are consistent, and are both steeper than the self-similar prediction of
$4/3$ (although not very significantly for the high-z data). At least some
of the larger scatter in the HIFLUGCS data is due to cooling cores in some
systems, as no correction for these was made \citep{rei02}. 

\begin{table*}
\centering
\scalebox{0.95}{
\begin{tabular}{llll} \hline 
\ML\ relation & $A$ ($10^{44}h_{70}^{-2}\ergps$) & $B$ & notes \\
\hline \hline

\multicolumn{4}{l}{\bf Properties within $R_{200(z)}$}\\

WARPS \ML\ & $2.76\pm0.81$ & $1.90\pm0.49$ & $0.6<z<1.0$, luminosities scaled for predicted evolution (Fig. \ref{c4fig_ML}).\\

Unscaled WARPS \ML\ & $13.8\pm5.4$ & $2.08\pm0.62$ & $0.6<z<1.0$, no scaling for predicted evolution (Fig. \ref{c4fig_ML}).\\

HIFLUGCS \ML\ & $3.40\pm0.34$ & $1.70\pm0.11$ & Local clusters assuming isothermality (Fig. \ref{c4fig_ML}).\\

\hline

\end{tabular}
}
\caption[Summary of the \ML\ relations discussed in \textsection
\ref{c4sect_ML}]{\label{c4tab_ML}Summary of the \ML\ relations discussed in
\textsection \ref{c4sect_ML}.} 
\end{table*}

\section{Discussion}
The normalisation of the WARPS high-redshift scaling relations are all
consistent with self-similar evolution of the local relations in a \LCDM
cosmology. We emphasise the importance of properly quantifying the effect
of assuming isothermality on the derived masses of high-redshift clusters
when investigating the evolution of the \MT\ relations. The \LT\ and
\MnT{g\ 2500(z)}\ relations are the most robust of the high-redshift
relations measured here, subject to the smallest systematic uncertainties
and extrapolations. In particular the combined V02 and WARPS high-z sample
enable the \LT\ relation to be measured with relatively small
uncertainties. The \ML\ relation meanwhile is predicted to show the
strongest evolution ($\propto E(z)^{-7/3}(\Dv(z)/\Dv(0))^{-7/6}$). The
consistency of these relations with the predicted self-similar evolution
provides strong evidence that that model is a good description of cluster
evolution out to $z\approx1$. 

A preliminary measurement of the high-redshift \MnT{2500(z)}\ relation in
\citet{mau03a} found that it was consistent with no evolution, but did not
strongly rule out self-similar evolution. This work improves on that
earlier study in several ways. The sample used is larger, and the latest
calibration was used in reanalysing the \Chandra\ data (see \textsection
\ref{c4sect_anal}). An important difference is that the masses are measured
within radii corresponding to redshift-dependent density contrasts,
estimated from the overdensity profile of each system. In the earlier
study, the masses were measured within a fraction of the virial radius
estimated from each cluster's temperature. The current study thus provides
a more reliable measurement of the evolution of the \MnT{2500(z)}\
relation. 

The slope of the combined WARPS and V02 \LT\ relation is consistent with
its low-redshift counterparts, and steeper than the slope of 2 predicted by
the self-similar model. This suggests that the same processes are
responsible for the steepening of both relations, and non-gravitational
processes have already influenced cluster properties by ($z\ge1$). A possible
interpretation of this is that the non-gravitational effects are important during
the early stages of clusters' lives, regardless of redshift, as the
clusters discussed here are generally relaxed in appearance, suggesting the
major part of their formation is complete. The high-redshift \ML\ relation
also supports this interpretation. 

The slopes of \MnT{200(z)}\ relation above $3\keV$ in both the WARPS and
S03 samples are consistent with the self-similar prediction of $3/2$. When
the cooler S03 systems are included, the slope steepens, which is
consistent with non-gravitational effects having a larger relative
contribution in low-mass systems. Observations of cooler ($<3\keV$)
clusters at high-redshift are required to test whether there is any
evolution in this effect. There is also a weak trend for the \MnT{2500(z)}\ and
\MnT{g\ 2500(z)}\ relations to be steeper than the \MnT{200(z)}\ and
\MnT{g\ 200(z)}\ relations, whose slopes agree more closely with the
self-similar slope. This trend is present, though not strongly significant,
in the high- and low-redshift relations. These results are consistent with
a scenario in which non-gravitational processes have a stronger effect in
the central regions of clusters, which is more noticeable in cooler
systems. 

Generally, the \MT\ relations show a self-similar slope above $3\keV$, the
\MgT\ relations are self-similar, or slightly steeper, and the \LT\
relations are steeper than self-similar prediction. This suggests that
although non-gravitational processes do not have a strong influence on the
dark matter in $>3\keV$ clusters, those processes can still have an
important effect on the gas in more massive systems. 

\subsection{Evolution of the \LT\ relation}
As we saw in the \textsection \ref{c4sect_LT}, the evolution of the \LT\
relation is consistent with the self-similar model described in
\textsection \ref{c4sect_theory}. The relatively large size of the combined
WARPS and V02 samples also enables alternative models of cluster evolution
that are used in the literature to be tested. As discussed in section
\textsection \ref{c4sect_theory}, it is popular to use a fixed,
redshift-independent density contrast to define the outer radius of
clusters. In this case, the \LT\ relation is given by
\begin{eqnarray}\label{c4eqn_LTa} 
L_{\Delta} E(z)^{-1} & \propto & kT^{B}.
\end{eqnarray}

Now the $\Delta(z)^{-1/2}$ part of the normalisation is no longer required,
and as this is an increasing function of $z$, the predicted \LT\ evolution
is smaller than that predicted by the $\Delta(z)$ model. The measured
luminosities are slightly larger when extrapolated to $R_\Delta$ instead of
$R_{\Delta(z)}$, which acts in the opposite sense, increasing the apparent
evolution. This evolution model was tested by comparing the combined WARPS
and V02 \LT\ relation, with luminosities extrapolated to $R_{200}$ (not
$R_{200(z)}$) and scaled by $E(z)^{-1}$, with the local AE99 relation. The
$E(z)^{-1}$ scaling was insufficient to reduce the luminosities to be
consistent with the local \LT\ relation, with $\chisq/\nu=51/22$. This
indicates that a redshift-dependent density contrast is required in the
self-similar model to reconcile the high-redshift and local \LT\ relations.

\begin{figure*}
\begin{center}
\scalebox{0.55}{\includegraphics*[angle=270]{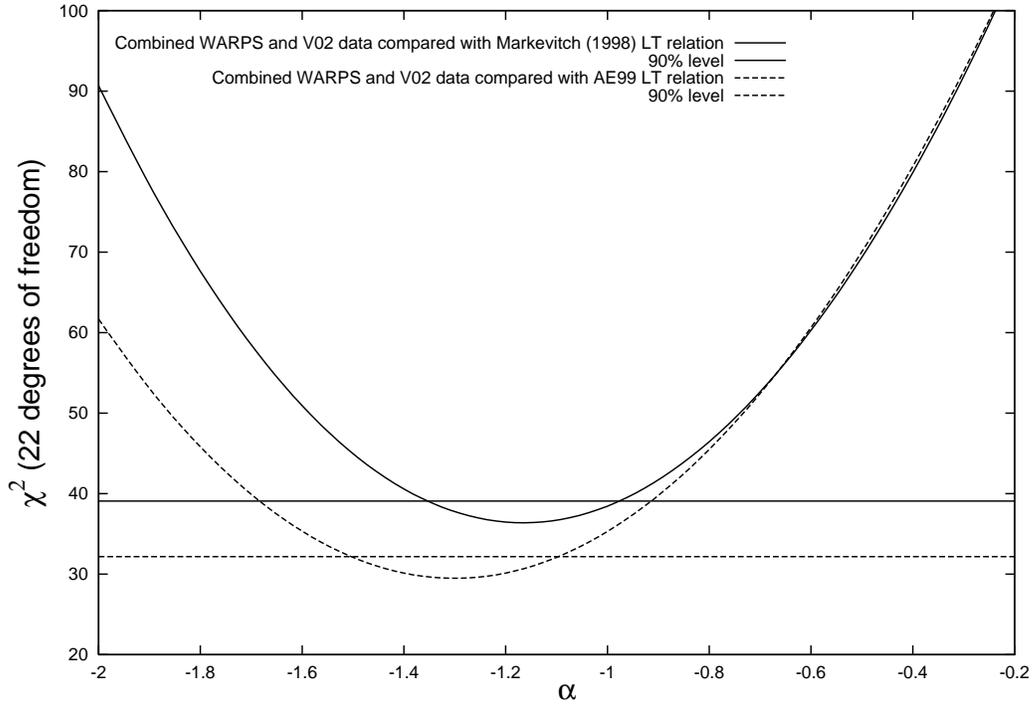}} \\
\caption[The $\chisq$ values for the comparison of the scaled high-redshift \LT\ data with the local relations.]{\label{c4fig_LTchi}The $\chisq$ values for the comparison of the scaled high-redshift WARPS and V02 \LT\ data with the local relations. The luminosities of the high-redshift systems were scaled by $(1+z)^{-\alpha}$ for different values of $\alpha$.}
\end{center}
\end{figure*}

A simple way of measuring the evolution of the \LT\ relation is to assume that it evolves as 
\begin{eqnarray}\label{c4eqn_LTb}
L_{\Delta} & \propto & (1+z)^\alpha kT^{B}.
\end{eqnarray}
V02 found $\alpha=1.5\pm0.3$ at $90\%$ confidence by comparing their
high-redshift data with the local \citet{mar98a} \LT\ relation. The
combined WARPS and V02 \LT\ data were used to make a similar
measurement. The luminosity of each cluster extrapolated to $R_{200}$ was
divided by $(1+z)^{\alpha}$, and the scaled data were compared with a local
relation. This process was performed for a range of values of $\alpha$, and
for both of the local AE99 and \citet{mar98a} \LT\ relations and the
results are shown in Fig. \ref{c4fig_LTchi}.

\begin{figure*}
\begin{center}
\scalebox{0.55}{\includegraphics*[angle=270]{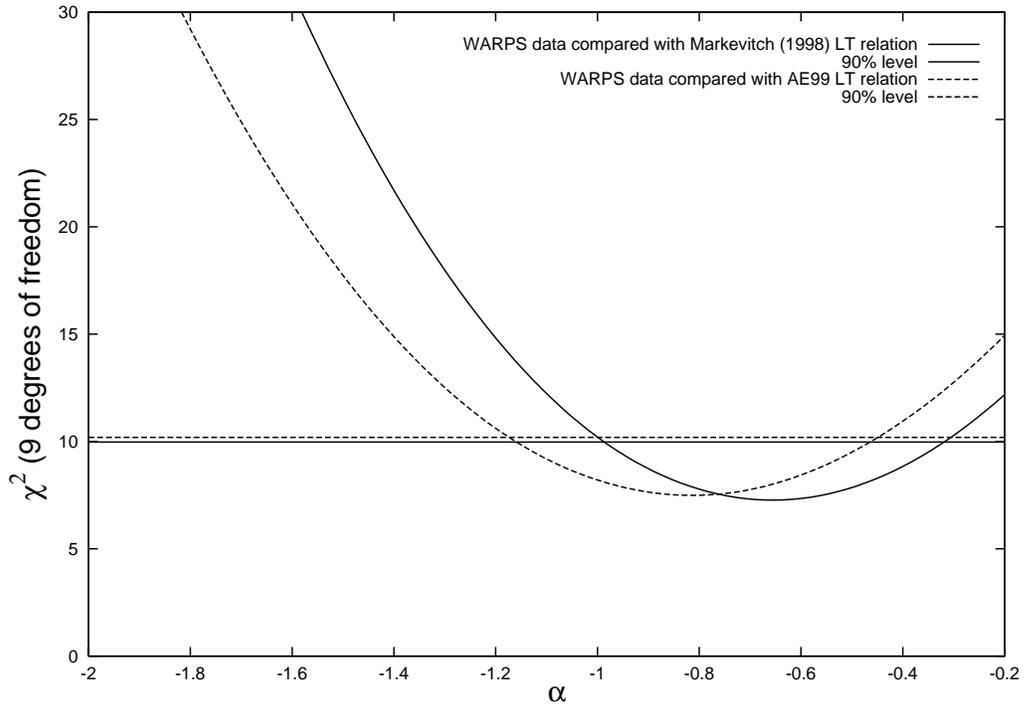}} \\
\caption[The $\chisq$ values for the comparison of the scaled high-redshift \LT\ data with the local relations.]{\label{c4fig_LTchiWARPS}The $\chisq$ values for the comparison of the scaled high-redshift WARPS \LT\ data alone with the local relations. The luminosities of the high-redshift systems were scaled by $(1+z)^{-\alpha}$ for different values of $\alpha$.}
\end{center}
\end{figure*}

We find the best agreement between the scaled high-redshift data and the
local AE99 relation, with $\alpha=1.3\pm0.2$ (at the $90\%$ level) and
$\chisq/\nu=23.0/22$. Using the \citet{mar98a} relation as a low-redshift
baseline results in a slightly lower value of $\alpha$ and a poorer fit. A
measurement of $\alpha$ independent of that of V02 was also performed by
using the WARPS sample alone. Compared with the AE99 local relation we find
$\alpha=0.8\pm0.4$ ($90\%$ level) while comparison with the \citet{mar98a}
relation gives $\alpha=0.7\pm0.4$ ($90\%$ level; see
Fig. \ref{c4fig_LTchiWARPS}. These measurements are marginally consistent
with the values of $\alpha=1.5\pm0.3$ ($68\%$ level) found by
\citet{lum04c} and $1.8\pm0.3$ ($68\%$ level) found by \citet{kot05} in
samples of high-redshift clusters observed with \XMM. We thus conclude that
$\alpha\approx1.2$ and that the range of values found is at least partially
due to the uncertainty in the exact form of the local \LT\ relation. 

While current the data cannot distinguish between a $L \propto
(1+z)^{\alpha}kT^B$ model and the full self-similar $L
E(z)^{-1}\Delta(z)^{-1/2} \propto kT^B$ model for the evolution of the \LT\
relation, the latter is more satisfactory because it is physically
motivated, and does not introduce an aditional free parameter ($\alpha$).

Recent theoretical work has attempted to include the effects of preheating and radiative cooling in the simple self-similar scaling relations. \citet{voi05} shows that introducing a cooling threshold $K_\mathrm{c}=T^{2/3}t(z)^{2/3}$, where gas with an entropy less than $K_\mathrm{c}$ radiates all of its thermal energy away within a Hubble time $t(z)$, produces a ``cooling threshold'' \LT\ relation of the form
\begin{eqnarray}\label{c4eqn_LTc}
L & \propto & kT^{2.5}E(z)^{-1}t(z)^{-1}.
\end{eqnarray}
\citet{voi05} also investigates the effect that including a modest initial amount of entropy in the gas before it is accreted onto a forming cluster has on the \LT\ relation. If the initial entropy level is chosen to match the observation that $K(0.1R_{200})\propto kT^{2/3}$ \citep{pon03}, then the resulting ``altered similarity'' \LT\ relation is
\begin{eqnarray}\label{c4eqn_LTd}
L & \propto & kT^{3}E(z)^{-3}t(z)^{-2}.
\end{eqnarray}

\begin{figure*}
\begin{center}
\scalebox{0.55}{\includegraphics*[angle=270]{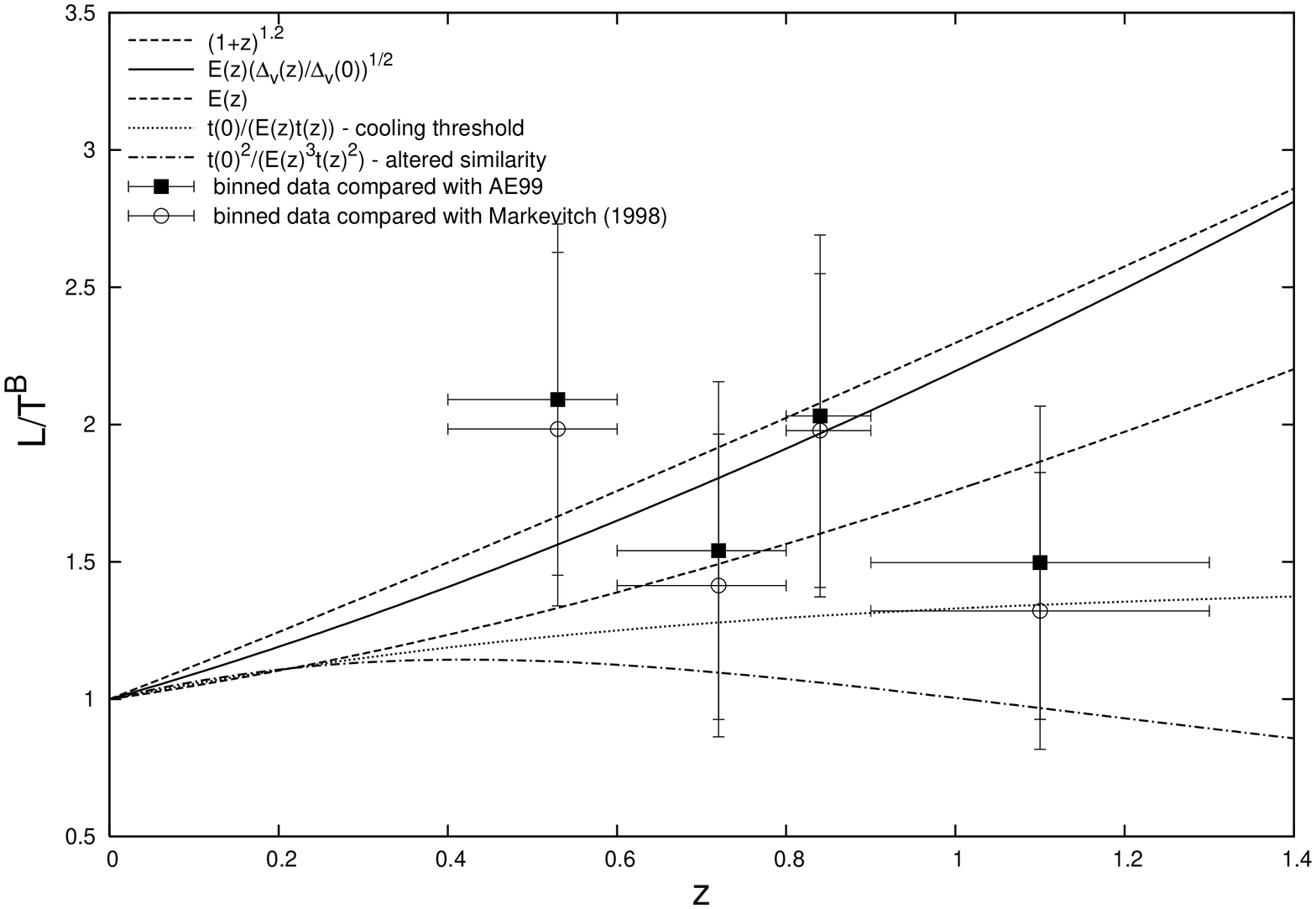}} \\
\caption[]{\label{fig_LTnormb}The evolution of the \LT\ relation normalisation predicted by different models. The data points are the observed evolution in the combined WARPS and V02 sample, in redshift bins, relative to different local \LT\ relation.}
\end{center}
\end{figure*}

Both of these modified self-similar models predict steeper \LT\ slopes
which are in better agreement with observations than the simple
self-similar slope of 2. The evolution of the \LT\ relation, however, is
much milder in the modified self-similar models than in the simple
self-similar model. Fig. \ref{fig_LTnormb} shows the evolution of the
normalisation of the \LT\ relation for a variety of models. The data points
show the observed evolution in the combined WARPS and V02 sample relative
to the AE99 and \citet{mar98a} local relations. The data are binned by
redshift, and the data points show the weighted mean and weighted standard
deviation for each bin. While dividing the data into redshift bins may be
pushing the limits of the current data, Fig. \ref{fig_LTnormb} serves to
illustrate the range of predictions for the evolution of the \LT\ relation.
The overall conclusion on the evolution of the \LT\ relation is that the
self-similar model is a good description of the evolution out to
$z\approx1$. Larger samples at high redshifts will enable more detailed
models to be tested, for example, indicating whether the evolution continues to
increase with redshift, or flattens out in line with the predictions of
the modified self-similar models.

\subsection{Continuous formation predictions}
The \MT\ relation defined in Equation \ref{c4eqn_MT} is derived under the
late-formation approximation, in which a cluster forms in a single
collapse, terminating with the system having just virialised at the
redshift of observation. Several authors \citep[\egc][]{lac93,voi00} have
derived \MT\ relations in a more realistic continuous-formation scenario,
in which clusters grow by accumulating much smaller virialised objects. In
a flat \LCDM cosmology, \citet{voi00} predicts 
\begin{eqnarray}\label{c4eqn_MTcf}
kT & = & (8.0\keV)\left(\frac{M}{10^{15}h_{100}^{-1}\Msol}\right)^{2/3}\frac{\xi_c(t)}{\xi_c(t_0)}
\end{eqnarray}
where $\xi_c(t)$ is the specific energy of a shell of matter which
collapses onto the cluster at a time $t$ (see the appendix of
\citealt{voi00} for details). In this model, the evolution of the
normalisation of the \MT\ relation is different from the late-formation
predictions.  

%Fig. \ref{c4fig_MTevol} shows the evolution of the \MT\ relation normalisation as a function of redshift in the late-formation and continuous-formation models.

%\begin{figure*}
%\begin{center}
%\scalebox{0.55}{\includegraphics*[angle=270]{mtnorm_bw.ps}} \\
%\caption{\label{c4fig_MTevol}Evolution of the normalisation of the \MT\ relation in late- and continuous-formation models.}
%\end{center}
%\end{figure*}

The continuous-formation model predicts that clusters of a given mass are
cooler than the late-formation model predictions. However, the difference
between the models is small at $z\approx1$, requiring measurements of the
\MT\ normalisation to a precision of $<2\%$ to distinguish between them,
and our results are consistent with either of these cluster-formation
scenarios. The agreement between our measurements of the evolution of the
\MT\ and \MgT\ relations, and these models indicates that the properties of
galaxy clusters reflect the properties of the universe at their redshift of
observation. In the late-formation approximation this is simply because
clusters formed at the redshift of observation, while in the
continuous-formation model, the accretion of matter onto clusters
continually realigns their properties with those of the evolving universe.

\subsection{Probing the redshift of virialisation}
If clusters form in a single collapse, then they may virialise at a
redshift ($z_\mathrm{v}$) which is larger than their redshift of
observation ($z_\mathrm{obs}$). Their properties would then reflect those
of the universe at an epoch earlier than $z_\mathrm{obs}$, and the
self-similar evolution which assumes $z_\mathrm{obs}=z_\mathrm{v}$ would
not be a good description of the high-redshift scaling relations. If one is
willing to assume that clusters obey self-similar evolution, then the
high-redshift data can be used to place interesting constraints on the mean
redshift of formation of the high-redshift sample
($\bar{z_\mathrm{vh}}$). If the mean redshift of formation of the local
sample is denoted as $\bar{z_\mathrm{vl}}$, then the ratio of the
normalisation of the low- and high-redshift \LT\ relations is given by
$[E(\bar{z_\mathrm{vh}})\Dv(\bar{z_\mathrm{vh}})^{1/2}]/[E(\bar{z_\mathrm{vl}})\Dv(\bar{z_\mathrm{vl}})^{1/2}]$
(\cf Equation \ref{c4eqn_LT}).

\begin{figure*}
\begin{center}
\scalebox{0.55}{\includegraphics*[angle=270]{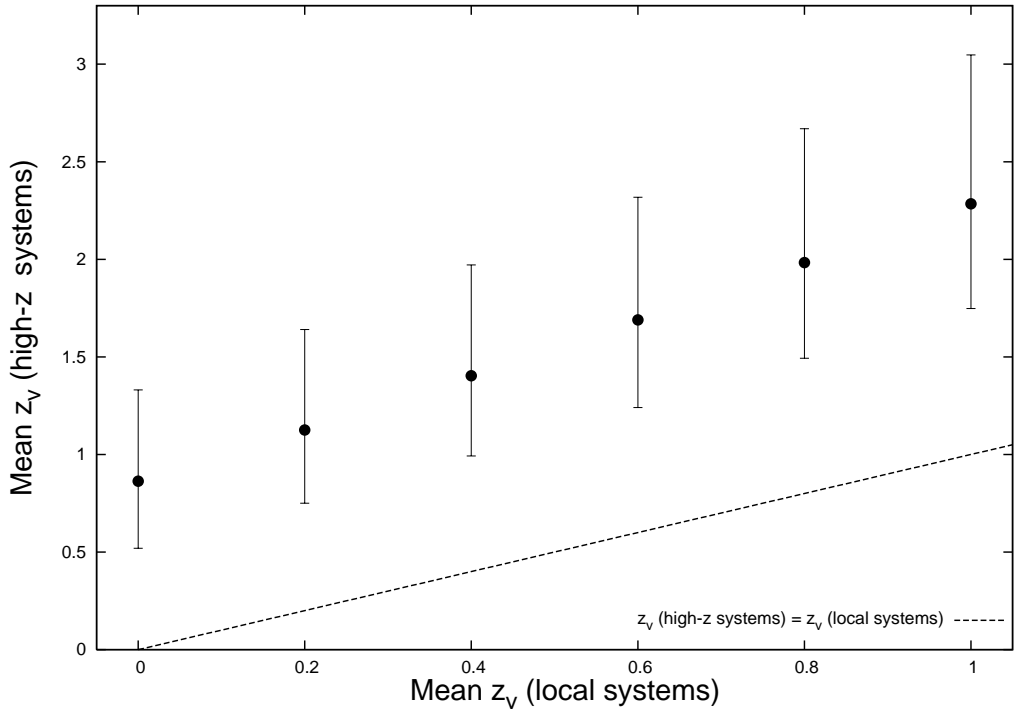}} \\
\caption{\label{c4fig_zform}Constraints on the mean redshift of virialisation of the combined WARPS and V02 high-redshift sample derived from the evolution of the \LT\ relation. The dotted line indicates a common redshift of virialisation of the high- and low-redshift samples. Errorbars are $99\%$.}
\end{center}
\end{figure*}

Under the assumptions given above, the ratio of the normalisations of the
high- and low-redshift \LT\ relations can be used to relate
$\bar{z_\mathrm{vl}}$ and $\bar{z_\mathrm{vh}}$. The high-redshift \LT\
relation thus allows $\bar{z_\mathrm{vh}}$ to be measured for an assumed
$\bar{z_\mathrm{vl}}$.  A relation with a slope fixed
at the local value of $2.88$ (AE99) was fit to the unscaled WARPS and V02
high-z \LT\ data, and the best-fit normalisation, along with the maximum
and minimum normalisations allowed by the data at the $99\%$ level were
found. These were then used to derive the constraints on
$\bar{z_\mathrm{vh}}$ plotted in Figure \ref{c4fig_zform}.

If the local or high-redshift clusters formed at a redshift other than
$z_\mathrm{obs}$, then the overdensity radii within which the luminosities
are measured would be inappropriate. However, as the overdensity radii are
all large and the surface-brightness is low in outer regions of the
clusters, the use of different radii has a negligible effect. To illustrate
this, if we consider a $6\keV$ cluster with $\beta=0.67$ and $\rc=100\kpc$
observed at $z=0$, we find $R_{200(z)}=2.0\Mpc$ for $z_\mathrm{v}=0$ and
$R_{200(z)}=1.6\Mpc$ for $z_v=1$. The luminosity of the cluster increases
by just $1\%$ between these two radii.

Figure \ref{c4fig_zform} shows that if the local clusters are assumed to
have virialised at $z=0$, then the high-redshift data are consistent with
$z_\mathrm{vh}=z_\mathrm{obs}$, with $\bar{z_\mathrm{vh}}<1.5$ at the
$99\%$ level. We strongly rule out a common redshift of virialisation for
the local and high-redshift samples, regardless of the redshift of
formation of the local sample. That is not to say that the high-redshift
clusters cannot evolve onto the local \LT\ relation, via mergers and
accretion resetting their properties to new values of $z_v$. These results
are simply indicating that the local clusters did not form fully
at the same high redshift as the distant clusters and then evolve
passively.

\subsection{Properties of the WARPS clusters}
The observed gas properties of the WARPS high-redshift sample are generally
consistent with those of local clusters. The mean surface-brightness
profile slope, $\bar{\beta}=0.66\pm0.05$ is very close to the canonical
value of $2/3$ \citep{jon84}. Also, the mean ratio of core radius to
``virial radius'' $\rc/R_{200(z)}=0.14\pm0.05$ agrees well with the values
found by \citet{san03b} for $>1\keV$ clusters. Furthermore, the gas-mass
fractions found in the high-redshift clusters agree with those found in
local clusters. The weighted mean \fgas\ at $R_{2500(z)}$ is
$0.069\pm0.012$, increasing to $0.11\pm0.02$ at $R_{200(z)}$ which agrees
with the values found by S03. These results indicate that the gas
distribution is the same in low- and high-redshift clusters.  

\begin{figure*}
\begin{center}
\scalebox{0.55}{\includegraphics*[angle=270]{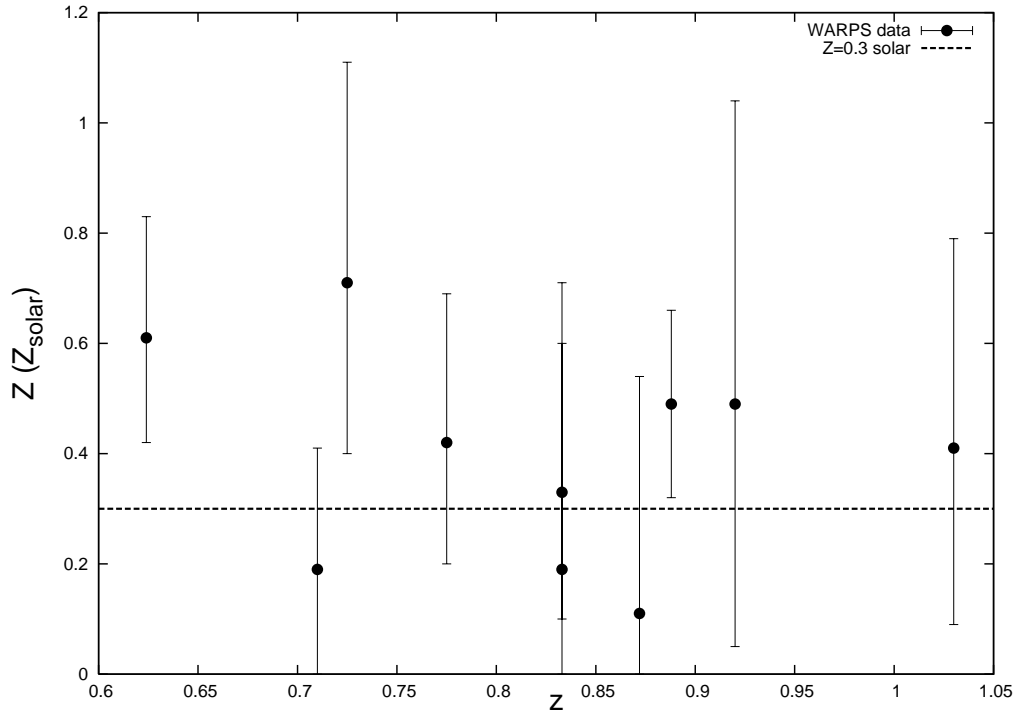}} \\
\caption{\label{c4fig_zZ}Cluster metallicity plotted against redshift for the WARPS high-redshift sample. Note that the poorly constrained high metallicity value measured in ClJ1559.1$+$6353 is not included in this plot.}
\end{center}
\end{figure*}

The metal abundances measured in the high-redshift systems are plotted
against redshift in Fig. \ref{c4fig_zZ}. The data (excluding
ClJ1559.1$+$6353 with its very high, poorly constrained measurement) are
consistent with the canonical value of $0.3\Zsol$ out to $z=1$; the
weighted mean of the values is $0.28\pm0.11\Zsol$. This is consistent with
the high-redshift of enrichment of the intra-cluster medium found by other
authors \citep[\egc][]{mus97a,toz03}. 

\section{Conclusions}
The overall picture provided by this study of the evolution of the cluster scaling relations is that within the statistical limits of the current data, the evolution of galaxy clusters out to $z\approx1$ is described well by the self-similar model. The large-scale properties of clusters are dominated by the density of the universe at the epoch at which they are observed.

\section{Acknowledgements}
We thank Alastair Sanderson for providing us with properties at different radii for the S03 sample. We are grateful to Stephen Helsdon for providing some of the software used for the regression line fitting. BJM was supported for the majority of this work by a Particle Physics and Astronomy Research Council (PPARC) postgraduate studentship. BJM is currently supported by NASA through Chandra Postdoctoral Fellowship Award Number PF4-50034 issued by the Chandra X-ray Observatory Center, which is operated by the Smithsonian Astrophysical Observatory for and on behalf of NASA under contract NAS8-03060. HE gratefully acknowledges financial support from NASA grant NAG 5-10085.

\bibliographystyle{mn2e}
\bibliography{clusters}

\end{document}